\newif\iffigs
 \figstrue
\documentclass[12pt]{article}
\usepackage{colordvi}
\usepackage{latexsym,amssymb}
\iffigs
    \usepackage{graphicx}
    \input{epsf}
\else
\fi
\newcommand{\ft}[2]{{\textstyle\frac{#1}{#2}}}
\def\tilde{\widetilde}

\newsavebox{\uuunit}
\sbox{\uuunit}
                             {\setlength{\unitlength}{0.825em}
                                      \begin{picture}(0.6,0.7)
                                                                  \thinlines
                                                                  \put(0,0){\line(1,0){0.5}}
                                                                  \put(0.15,0){\line(0,1){0.7}}
                                                                  \put(0.35,0){\line(0,1){0.8}}
                                                                 \multiput(0.3,0.8)(-0.04,-0.02){10}{\rule{0.5pt}{0.5pt}}
                                      \end {picture}}

\makeatletter \@addtoreset{equation}{section} \makeatother

\def\bfone{\relax{\rm 1\kern-.35em 1}}

\def\tilde{\widetilde}
\def\hat{\widehat}

\begin{document}
\begin{titlepage}
\begin{flushright}
DISTA-2007 \\
hep-th/yymmnnn
\end{flushright}
\vskip 1.5cm
\begin{center}
{{\LARGE \bf Pure Spinor Superstrings on Generic\\}
\vskip 0.1cm
{\LARGE \bf Type IIA Supergravity Backgrounds$^\dagger$}}
 \vfill
\vskip 0.5cm
{\large  R. D'Auria$^1$, P. Fr{\'e}$^2$, P. A. Grassi$^3$, and M. Trigiante$^1$}
\vfill {
$^1$ Dipartimento di Fisica Politecnico di Torino,\\ C.so Duca degli
Abruzzi, 24,
I-10129 Torino, Italy, \\
$^2$ Dipartimento di Fisica Teorica, Universit{\`a} di Torino, \\
$\&$ INFN -
Sezione di Torino\\
via P. Giuria 1, I-10125 Torino, Italy\\
\vskip .3cm
$^3$
{ DISTA, Universit\`a del Piemonte Orientale, }\\
{ Via Bellini 25/G,  Alessandria, 15100, Italy
$\&$ INFN - Sezione di
Torino
\vskip 1cm
}}
\end{center}
\vfill
\begin{abstract}
We derive the Free Differential Algebra for type IIA supergravity in 10 dimensions in the
string frame. We provide all fermionic terms for all curvatures. We derive the Green-Schwarz
sigma model for type IIA superstring based on the FDA construction and we check its
invariance under $\kappa$-symmetry. Finally, we derive the pure spinor sigma model
and we check the BRST invariance. The present derivation has the advantage that the resulting
sigma model is constructed in terms of the superfields appearing in the FDA and therefore one
can directly relate a supergravity background with the corresponding sigma model. The complete
explicit form of the BRST transformations is given and some new pure spinor constraints are obtained.
Finally, the explicit form of the action is given.

\end{abstract}
\vfill
\vspace{1.5cm}
\vspace{2mm} \vfill \hrule width 3.cm {\footnotesize $^ \dagger $
This work is supported in part by the European Union RTN contract
MRTN-CT-2004-005104 and by the Italian Ministry of University (MIUR) under
contracts PRIN 2005-024045 and PRIN 2005-023102}
\end{titlepage}
\tableofcontents
\newpage
\section{Introduction}
\label{introibo}

The pure spinor formulation of superstrings is a new formalism \cite{Berkovits:2000fe} which
powerfully uses the advantages of the RNS formulation and those of the GS formalism. In particular,
the purpose of its creation was to provide a set-up  where the RR fields (appearing in the
spectrum of superstrings) could be treated on the same footing as the NSNS ones. This equal-footing
treatment of the bosonic massless modes of superstrings is realized in
every formulation of supergravity (in components, in superspace or, using rheonomic approach).
Therefore it would be convenient also for the pure spinor sigma model. This means that the couplings
of the worldsheet fields with the RR backgrounds must be very similar to the coupling  with the NSNS fields. This is indeed achieved in the pure spinor formulation.

Dealing with the complete supergravity multiplet and with its
non-linear self-interactions requires a full-fledged formulation
of pure spinor superstrings on arbitrary (on-shell) background.
This has been achieved in the fundamental work
\cite{Berkovits:2001ue} where a generic sigma model, respecting
the requirements of super-Poincar\'e invariance (both on the
worldsheet as well as in the target space) and with the correct
quantum numbers has been constructed. Consequently, according to
the  formulation, two BRST currents and their charges are
provided. Thus, imposing the nilpotency of these BRST charges
(which is equivalent to the closure of the constraint algebra) and
the holomorphicity of their currents  (which is equivalent to the
invariance of the action), the authors derived the supergravity
equations of motion in the form of superspace constraints. The
main input in \cite{Berkovits:2001ue}  is the requirement of the
constraints on the ghost fields
$$
\bar\lambda_1 \Gamma^m \lambda_1=0\,,
~~~~~~~~~
\bar\lambda_2 \Gamma^m \lambda_2=0\,.
$$
Here
$\bar\lambda_i = \lambda^T_i C$ with $C$ is the charge conjugation matrix. The index
 $i$ stands for the right- or the left-mover pure spinors whose chirality is decided by choosing
 either IIA or IIB. These constraints  are necessary  for the nilpotency of the
 BRST charge in the flat limit and they are essential to establish the correct number of degrees
 of freedom. Therefore, they have been imposed also for the interacting
 sigma model on generic backgrounds. Doing that, the emerging superspace constraints
 have a complicated and unconventional relation with
 the standard description of supergravity.  Yet, in \cite{Berkovits:2001ue} it is argued
 how, using Weyl superspace \cite{Howe:1997rf}, one can relate
 the supergravity constraints from the pure spinor formulation with those
 given in \cite{Howe:1983sra,Carr:1986tk}.
 To be more explicit, the connection between a more conventional setting and the
 pure spinor formulation is obtained by a Weyl transformation involving the dilatino followed
 by a Poincar\'e transformation needed to reabsorb some additional terms in the variation
 of the gravitino fields.  Thus, the conclusion is that, insisting on very simple pure
 spinor constraints, the ensuing supergravity parametrization in superspace turns out to be
 rather obscure. This fails to provide a practical and an effective algorithm to deduce
 the pure spinor sigma model starting from a given supergravity background.

 Let us invert the path. The old path goes  from pure spinor constraints to the sigma model and
 yields the supergravity constraints as a by-product.  The new path goes from the geometrical
 formulation of  supergravity to the  pure spinor
 sigma model. Indeed, we decide
 to start from a convenient description of supergravity and
 deduce the constraints and the conditions under which a pure spinor sigma model
 can exist.

\par
For these reasons it is highly desirable to have a formulation
of the pure spinor sigma models in which the pure spinor constraints,
the BRST operator and the entire set up follow from background
supergravity as it happens for the $\kappa$-symmetric
actions.
\par
Such a formulation is presented in this paper. Previous work in
this direction was accomplished in
\cite{Fre':2006es,Fre:2007xy,Fre:2008qw}, where such ideas were
applied to the case of M-theory and of the M2-brane. Here we focus
on type II superstrings and in particular on the type IIA case.
This is not a random choice but it is motivated by precise
reasons. Our ultimate goal is three-fold, since we want to show
that:
\begin{description}
  \item[1] The pure string BRST invariant $\sigma$-model  can be
constructed on arbitrary supergravity backgrounds;
  \item[2] The structure of the BRST algebra, the form of the pure
  spinor constraints and  the
  $2$-dimensional action can be algorithmically derived from
  supergravity and its Free Differential Algebra;
  \item[3] The explicit form of the $\sigma$-model action obtained in
  this way is of immediate practical use for dealing with
  backgrounds characterized by less than maximal supersymmetry, like
  $\mathrm{AdS} \times \mathrm{M}$ supergravity solutions where
  $\mathrm{M}$ is not a sphere.
\end{description}
As we already discussed in \cite{Fre:2007xy}, issue 3)
consists of solving the supergravity problem of
\textit{supergauge completion}. This means the explicit integration in
superspace of the rheonomic conditions which are first order differential
equations in the Grassmann $\theta$-variables. Such integration is
just a brute-force matter (see for example the application to super-Yang-Mills in 10d
\cite{Grassi:2004ih}), being a priori guaranteed by the
fulfillment of Bianchi identities and, it can be quite cumbersome in
general situations. In the directions of those $\theta$-variables
that correspond to supersymmetries preserved by the chosen
background, the integration is automatically performed by the use of
Maurer-Cartan superforms of the superisometry algebra
(for instance $\mathrm{SU(2,2|1)}$ in the case of the
$\mathrm{AdS}_5 \times \mathrm{T^{(1,1)}}$ compactification of type
IIB supergravity\cite{Romans:1984an,Klebanov:1998hh,Ceresole:1999zs} or
$\mathrm{Osp(6|4)}$ in the case of the $\mathrm{AdS}_4 \times \mathbb{P}^3$
compactification of the type IIA theory \cite{Nilsson:1984bj}).
In the other directions, namely those along the $\theta$'s associated with broken
supersymmetries, the integration of the rheonomic conditions might
be involved. Hence, in order to explore the structure of the
\textit{supergauge completion} it is  desirable to have the
minimal possible amount of broken thetas. Among the possible
compactifications, one case is
the $\mathrm{AdS}_4 \times \mathbb{P}^3$ background.
There the preserved thetas are 24 and
the broken ones just 8, and they are arranged into an $\mathrm{O(2)}$ doublet of $\mathrm{D=4}$
spinors leading to the hope that the corresponding sigma model as a nice and insightful description.
It is therefore in such perspective we began to focus on
the type IIA case rather than on the type IIB one which will
follow \cite{noiIIB}.
\par
A second technical reason for this strategy will be clear to the
reader. In order to carry through our programme, the formulation
of supergravity which is required is in the \textit{string frame}
rather than that in the \textit{Einstein frame}. Although the two
formulations are simply related by a field redefinition, the
implementation of such a change of variables in the rheonomic
solution of the Free Differential Algebra Bianchi Identities is so
cumbersome that it turns out to be more convenient to redo the
construction of supergravity directly in the new frame. In view of
this we can say that neither the type IIA nor the type IIB theory
were available in the rheonomic framework and in the string frame
when we started the present work. Indeed the rheonomic type IIA
theory was never constructed, while the rheonomic type IIB case
was constructed by Castellani and Pesando in the \textit{Einstein
frame} \cite{castella2b,igorleo}.  The transition to the string
frame is even more elaborate in the  IIB case than in the IIA one,
due to the $\mathrm{SU(1,1)}$ covariance of the IIB theory, which
is made manifest only in the Einstein frame.
\par
Having clarified our motivations, let us summarize the structure
of the paper:

\begin{enumerate}
\item As already recalled above, the algebraic structure underlying any higher dimensional supergravity theory
is a Free Differential Algebra (FDA) \cite{fredauria11,comments}.
This latter is a categorical extension of a (super) Lie algebra
determined by the Chevalley cohomology of the latter
\cite{sullivan};
\item Given the FDA one considers its Bianchi identities and
constructs the unique rheonomic parametrization of the FDA
curvatures. Rheonomy is a universal principle of analiticity in
superspace \cite{castdauriafre} which requires that the fermionic
components of the FDA curvatures should be linear functions of
their bosonic ones. Rheonomy encodes in one single principle the
construction of both field equations and supersymmetry
transformation rules for any supergravity. Indeed field equations
follow as integrability conditions of the rheonomic
parametrization of curvatures. The flow chart for the construction
of classical supergravities was for instance recently presented in
\cite{Fre:2008qw};
\item Consider then the FDA appropriate to the supergravity under
investigation and the rheonomic parametrization of its curvatures;
\item Perform the ghost-form extension of the classical FDA
according to the principle introduced by Anselmi and Fr\'e in \cite{Anselmi:1992tj}, which generalizes
ideas previously introduced by Baulieu\cite{Baulieu:1986dp} namely:

{\it The BRST algebra is provided by replacing, in the rheonomic
parametrization of the classical supergravity curvatures, each
differential form with its extended ghost-form counterpart while
keeping the curvature components untouched. Thus one obtains the
rheonomic parametrization of the ghost--extended curvatures, whose
formal definition is identical with that of the classical
curvatures with the replacements:}
\begin{equation}
  \begin{array}{ccc}
    d & \mapsto & d+\mathcal{S} \\
    \Omega^{[n]} & \mapsto & \sum_{p=0}^n \, \Omega^{[n-p,p]} \
  \end{array}
\label{orletto1}
\end{equation}
where ${\cal S}$ is the BRST differential and $\Omega^{[n-p,p]}$ is a
ghost form with form degree $n-p$ and  ghost number $p$.

 In this
way one has the ordinary (unconstrained) BRST algebra of
supergravity;
\item Set to zero all the bosonic ghosts. This defines a
constrained BRST algebra and for consistency a certain set of
{\it pure spinor} constraints. The correct constraints are the
  projection onto the world-sheet (brane world volume) of these
  constraints.;

  \item Verify that the pure spinor constraints can be solved in
  terms of as many independent degrees of freedom as it is required
  for a conformal theory in d=2 in the case of superstrings with vanishing central
  charge;
  \item Introduce the appropriate antighosts and Lagrange multiplier
  field and construct the BRST invariant quantum action.
\end{enumerate}

The whole procedure can be summarized as follows:

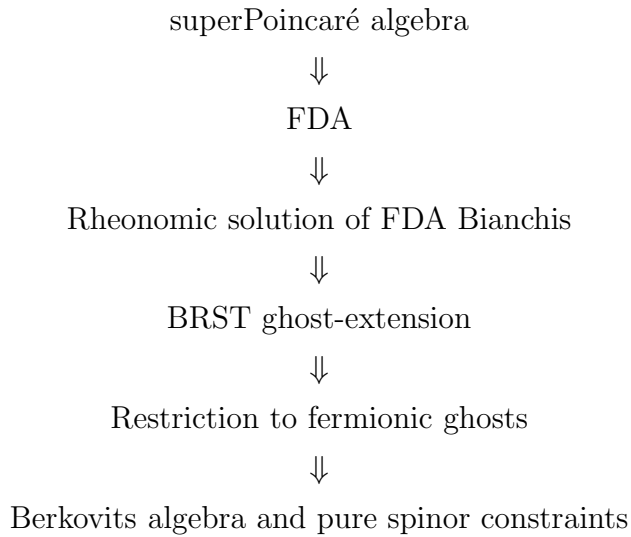
\begin{table}[!hbt]
  \centering
  $$
  \begin{array}{c}
    \mbox{superPoincar\'e algebra} \\
    \Downarrow  \\
    \mbox{FDA} \\
    \Downarrow \\
    \mbox{Rheonomic solution of FDA Bianchis} \\
    \Downarrow \\
    \mbox{BRST ghost-extension} \\
    \Downarrow \\
    \mbox{Restriction to fermionic ghosts} \\
     \Downarrow \\
\mbox{Berkovits algebra and pure spinor constraints}\\
  \end{array}
  $$
\end{table}
In this way we determine a path from the superPoincar\'e algebra
to the Berkovits BRST algebra on the fields of non negative
ghost-number (see the above flowchart). As we pointed out in
\cite{Fre:2008qw} the inclusion of the extra fields with negative
ghost number (the antighosts) requires more explanation since it
is not a standard gauge-fixing procedure but it is  obviously
essential for the construction of the $\sigma$-model action.
\par
We explicitly show how to realize the last steps of the
construction in the case of the type IIA theory and we emphasize
that they are just possible because of some very special features
of the rheonomic solution of the FDA Bianchi identities which are
displayed by its \textit{string frame} formulation and are instead
absent in the \textit{Einstein frame}.
\par
The result of our construction is an explicit expression of the
pure spinor BRST invariant action of type IIA superstrings holding
true on any supergravity background, irrespectively of the number
of supersymmetries it preserves. As a by-product of the
construction we have also the emission vertices  for all the
supergravity fields, both fermionic and bosonic, both of the
Neveu-Schwarz and of the Ramond Ramond sectors.
\par
Our paper is organized as follows: In sec. 2, we discuss the
formulation of supergravity using the Free Differential Algebra in
the string frame. We compute the complete parametrization of the
fermionic and bosonic curvatures, including the 3 and 4-fermion
terms. In sec. 3, we construct the Green-Schwarz sigma model for
type IIA superstring using the FDA and we discuss the background
independence. In sec. 4, we provide the pure spinor formulation of
superstring based on the BRST transformations obtained from the
FDA. In appendices we supplement the main text with some detail of
the derivation and the conventions.


\section{Type IIA Supergravity and its FDA}
\label{sugra2agen}
Free Differential Algebras (FDA) are a natural categorical extension of the
notion of Lie algebra and constitute the natural mathematical
environment for the description of the algebraic structure of higher
dimensional supergravity theory, hence also of string theory. The
reason is the ubiquitous presence in the spectrum of
string/supergravity theory of antisymmetric gauge fields ($p$--forms)
of rank greater than one.
\par
FDA.s were independently discovered in Mathematics by Sullivan \cite{sullivan}
and in Physics by two of the authors of this paper (R. D'Auria and P. Fr\'e)
\cite{fredauria11}. The original name given to this algebraic structure
by D'Auria and Fr\'e was that of \textit{Cartan Integrable Systems}.
Later, recognizing the conceptual identity of this supersymmetric construction with the pure
bosonic constructions considered by Sullivan, we also turned to its naming FDA which has by now become
generally accepted.
\par
Let us also recall that the classification and the explicit construction of FDA.s relies on
two structural theorems by Sullivan
showing how all possible such algebras are \textit{cohomological extensions} of normal Lie algebras or
superalgebras (for a recent and short review of these concepts just adapted to our purposes
see \cite{Fre:2008qw}).
\par
The Free Differential algebra of type IIA supergravity in
$\mathrm{D=10}$ can be  obtained by dimensional reduction on a
circle $\mathbb{S}^1$ from the FDA of the $\mathrm{D=11}$
supergravity \cite{cremmerjulia,CFD11}. Although straightforward
this construction was never shown in the literature and it is
quite lengthy and laborious. For this reason in appendix
\ref{reduczia} we sketch the main steps of such a derivation.
Furthermore, as we explain extensively in the sequel, our main
target is the rheonomic parametrization of the FDA curvatures in
the string frame and not in the Einstein frame. Hence in the
quoted appendix we develop a mixed strategy to obtain our goal. We
begin by constructing the rheonomic parametrization of the bosonic
curvatures in the Einstein frame using dimensional reduction from
D=11. In this way we also obtain the bosonic field equations of
type IIA supergravity from dimensional reduction which is an
easier task than deriving them from the Bianchi identities or from
the construction of the  $D=10$ action. Next we perform a Weyl
transformation to the string frame which changes the bosonic field
equations only by an easy rescaling and once we have obtained the
rheonomic parametrizations of the FDA curvatures in the string
frame we directly determine the rheonomic parametrization of the
fermionic curvatures in that frame from the analysis of the
Bianchi identities.
\par
All the above mentioned steps are discussed in the appendix. In
the main text, we simply present the final result, namely type IIA
supergravity in the string frame.
\subsection{Type IIA FDA in the string frame}
\par
The field content of
type IIA supergravity is given in table \ref{tabella2}.
{
\begin{table}
  \centering
  \caption{\textbf{Field content of type IIA supergravity}.}
 \label{tabella2}
 {\small
$$
  \vbox{
  \offinterlineskip
  \halign
    {&\vrule# &\strut\hfil #\hfil &\vrule# &\hfil #\hfil &\vrule#
     &\hfil #\hfil &\vrule# &\hfil #\hfil  &\vrule# \cr
   \noalign{\hrule}
      height5pt
      &\omit& &\omit& &\omit& &\omit&
      \cr
      &\hbox to4 true cm{\hfill Form/degree \hfill}&
      &\hbox to4 true cm{\hfill string sector \hfill}&
      &\hbox to4 true cm{\hfill $\mathrm{SO(1,9)}$-rep/Chirality \hfill}&
      &\hbox to4 true cm{\hfill \vbox {\hbox{superstring}
       \hbox{zero modes}} \hfill}&
      \cr
      height5pt
      &\omit& &\omit& &\omit& &\omit&
      \cr
    \noalign{\hrule}
      height5pt
      &\omit& &\omit& &\omit& &\omit&
      \cr
      &$V^a$ - $[1]$& &NS-NS& &$(2,0,0,0,0)$& &\mbox{graviton $h_{\mu\nu}$}& \cr
      height10pt
      &\omit& &\omit& &\omit& &\omit&
      \cr
      height10pt
      &\omit& &\omit& &\omit& &\omit&
      \cr
      &$\psi_{R}$ - $[1]$& &R-NS& & $(\ft 32,\ft 12,\ft 12,\ft 12,\ft 12) $ - right &
       &$\mbox{ gravitino $\psi_{R\mu}$}$  & \cr
      height10pt
      &\omit& &\omit& &\omit& &\omit&
      \cr
      &$\psi_{L}$ - $[1]$& &NS-R& & $(\ft 32,\ft 12,\ft 12,\ft 12,\ft 12) $ - left &
       &$\mbox{ gravitino $\psi_{L\mu}$}$  & \cr
      height10pt
      &\omit& &\omit& &\omit& &\omit&
      \cr
      &$\mathbf{B}^{[2]}$ - $[2]$& &NS-NS& &$(1,1,0,0,0)$ & & Kalb-Ramond &\cr
      height10pt
      &\omit& &\omit& &\omit& &\omit&
      \cr
      &$\mathbf{C}^{[1]}$ - $[1]$& &R-R& &$(1,0,0,0,0)$& &R-R $1$-form& \cr
      height10pt
      &\omit& &\omit& &\omit& &\omit&
      \cr
      &$\mathbf{C}^{[3]}$ - $[3]$& &R-R& &$(1,1,1,0,0)$& &R-R $3$-form& \cr
      height10pt
      &\omit& &\omit& &\omit& &\omit&
      \cr
      &$\chi_R$& &NS-R& &$(\ft 12,\ft 12,\ft 12,\ft 12,\ft 12) $- right& &$\mbox{dilatino right}$& \cr
      height10pt
      &\omit& &\omit& &\omit& &\omit&
      \cr
      height10pt
      &\omit& &\omit& &\omit& &\omit&
      \cr
      &$\chi_L$& &R-NS& &$(\ft 12,\ft 12,\ft 12,\ft 12,\ft 12) $- left& &$\mbox{dilatino left}$& \cr
      height10pt
      &\omit& &\omit& &\omit& &\omit&
      \cr
      &$\varphi$& &NS-NS& &$(0,0,0,0,0)$ & &$\mbox{dilaton}$& \cr
      height5pt
      &\omit& &\omit& &\omit& &\omit&
      \cr
    \noalign{\hrule} }
  }
$$
}
  \label{fieldcont3}
\end{table}
} This field content corresponds to the basic forms of a specific
Free Differential Algebra including the $0$--form items entering
the rheonomic parametrizations of its curvatures.
\par
The starting point is, as usual, the superPoincar\'e algebra. In
$\mathrm{D=10}$ we have two superPoincar\'e algebras with $32$
supercharges, the type IIA and the type IIB. The Maurer Cartan
description of the type IIA superalgebra is obtained by setting to
zero the following curvatures:
\paragraph{Type IIA superPoicar\'e algebra in the string frame}
\begin{eqnarray}
R^{ {ab}} & \equiv & d\omega^{ {ab}} \, - \, \omega^{ {ac}}\, \wedge \,
\omega^{ {cb}}\label{2acurva}\\
T^{ {a}} & \equiv & \mathcal{D} \, V^{ {a}} \, - \,
{\rm i} \, \ft 12 \left(\overline{\psi}_L \, \wedge \,\Gamma^{ {a}} \, \psi_L \, + \,
\overline{\psi}_R \, \wedge \,\Gamma^{ {a}} \, \psi_R \right)
\label{2atorsio}\\
\rho_{L,R} & \equiv & \mathcal{D}\psi_{L,R} \, \equiv \, d\psi_{L,R} \, - \,
\ft 14 \omega^{ {ab}} \, \wedge \, \Gamma_{ {ab}} \,\psi_{L,R} \label{a2curlgrav}\\
\mathbf{G}^{[2]} & \equiv & d\mathbf{C}^{[1]} \, + \, \exp\left[ - \, \varphi \right]
\, \overline{\psi}_R \, \wedge \, \psi_L \label{F2defi}\\
\mathbf{f}^{[1]} & \equiv & d\varphi \label{dilacurva}\\
\nabla \chi_{L/R} &\equiv& \, d\chi_{L,R} \, - \,
\ft 14 \omega^{ {ab}} \, \wedge \, \Gamma_{ {ab}} \,\chi_{L,R} \label{a2curldil}
\end{eqnarray}
where the $0$--form dilaton $\varphi$ appearing in eq. (\ref{F2defi})
introduces a mobile coupling constant. Furthermore,
$V^{ {a}}, \omega^{ {ab}}$ denote the vielbein and
the spin connection $1$-forms, respectively, while the two fermionic $1$-forms
$\psi_{L/R}$ are Majorana-Weyl spinors of opposite chirality:
\begin{equation}
  \Gamma_{11} \, \psi_{L/R} \, = \, \pm \, \psi_{L/R}
\label{psiLR}
\end{equation}
The flat metric $\eta_{ {ab}}\, = \, \mbox{diag}(+,-,\dots,-)$
is the mostly minus one and $\Gamma_{11}$ is hermitian and
squares to the the identity $\Gamma_{11}^2=\mathbf{1}$.
\par
Setting
$R^{ {ab}}=T^{ {a}}=\mathbf{G}^{[2]}=\mathbf{f}^{[1]}=0$
one obtains the Maurer Cartan equations of a superalgebra where the
spinor charges, $Q_{L,R}$ dual to the spinor $1$-forms $\psi_{L,R}$
not only anticommute to the translations $P_a$ but also to a central
charge $Z$ dual to the (Ramond Ramond) $1$-form $\mathbf{C}^{[1]}$.
\par
According to Sullivan's second theorem the FDA extension of the
above superalgebra is dictated by its cohomology. In a first step
one finds that there exists a cohomology class of degree three
which motivates the introduction of a new $2$-form generator
$\mathbf{B}^{[2]}$ which in the superstring interpretation is just
the Kalb--Ramond field. Considering then the cohomology of the
FDA-extended algebra one finds a degree four cohomology class
which motivates the introduction of a $3$--form generator
$\mathbf{C}^{[3]}$. In the superstring interpretation, this is
just the second R-R field, the first being the gauge field
$\mathbf{C}^{[1]}$. Altogether the complete type IIA FDA is
obtained by adjoining the following curvatures to those already
introduced:
\paragraph{The FDA extension of the type IIA superalgebra in
the string frame}
\begin{eqnarray}
\mathbf{H}^{[3]} & = & d\mathbf{B}^{[2]} \, + \, {\rm i} \,
\left(\overline{\psi}_L \, \wedge \,
\Gamma_{ {a}} \, \psi_L \,
- \, \overline{\psi}_R \, \wedge \,\Gamma_{ {a}} \,
\psi_R \right) \, \wedge \, V^{ {a}}
\label{KRcurva}\\
\mathbf{G}^{[4]} & =
& d\mathbf{C}^{[3]} \, + \, \mathbf{B}^{[2]} \, \wedge \,
d\mathbf{C}^{[1]}\, \nonumber\\
&& - \, \ft 12 \, \exp\left [- \, \varphi \right]\,\left(\overline{\psi}_L \, \wedge \,
\Gamma_{ {ab}} \, \psi_R \,
+ \, \overline{\psi}_R \, \wedge \,\Gamma_{ {ab}} \, \psi_L
\right)\, \wedge \, V^{ {a}} \, \wedge \, V^{ {b}} \label{Gcurva}
\end{eqnarray}
Equations (\ref{2acurva}-\ref{dilacurva}) together with eq.s
(\ref{KRcurva}-\ref{Gcurva}) provide the complete definition of the
type IIA Free Differential Algebra.
\par
The next task is that of writing the Bianchi identities and construct
their rheonomic solution.
\paragraph{The Bianchi identities}
The curvature definitions listed above lead immediately to the
following Bianchi identities which we write, already under the assumption that the torsion is zero
$T^{ {a}}=0$:
\begin{eqnarray}
0 & = & \mathcal{D}\, R^{ {ab}} \label{LorentzBianchi}\\
0 & = & R^{ {ab}} \, \wedge \, V_{ {b}} \, - \,
{\rm i} \left(\overline{\psi}_L \, \wedge \,
\Gamma^{ {a}} \rho_L \, + \, \overline{\psi}_R \, \wedge
\, \Gamma^{ {a}} \rho_R
\right)\label{TorsionBianchi}\\
0 & = & \mathcal{D}\, \rho_{L/R} \, + \, \ft 14 \, R^{ {ab}}
\, \wedge \, \Gamma_{ {ab}} \, \psi_{L/R}\label{gravitinoBianchi}\\
0 & = & d\,\mathbf{G}^{[2]} \, + \, \mathbf{f}^{[1]} \, \wedge\,
\exp[-\varphi] \, \overline{\psi}_R \, \wedge \psi_L\, + \,
\exp[-\varphi] \, \left( \overline{\psi}_R \, \wedge \,
\rho_L-\overline{\psi}_L \, \wedge \,
\rho_R\right)\label{G2Bianchi}\\
0 & = & d\mathbf{f}^{[1]}\label{dilatonBianchi}\\
0 & = & d\,\mathbf{H}^{[3]} \, + \,
2\,{\rm i} \,  \left(\overline{\psi}_L \, \wedge \, \Gamma_{ {a}} \, \rho_L \, - \,
\overline{\psi}_R \, \wedge \, \Gamma_{ {a}}\, \rho_R
\right)\, \wedge \, V^{ {a}}\label{H3Bianchi}\\
0 & = & d\,\mathbf{G}^{[4]}\,  - \, \mathbf{H}^{[3]} \, \wedge \, \mathbf{G}^{[2]}
\, + \, {\rm i} \,
\left(\overline{\psi}_L \, \wedge \,
\Gamma_{ {a}} \, \psi_L \,
- \, \overline{\psi}_R \, \wedge \,\Gamma_{ {a}} \,
\psi_R \right) \, \wedge \, V^{ {a}} \, \wedge \,
\mathbf{G}^{[2]}\nonumber\\
&&+ \, \mathbf{H}^{[3]} \, \wedge \,
\exp\left[ - \, \varphi \right]
\, \overline{\psi}_R \, \wedge \, \psi_L\nonumber\\
&& - \, \ft 12 \, \mathbf{f}^{[1]} \,  \wedge \,
\exp\left[ - \, \varphi \right]\, \left(\overline{\psi}_L \, \wedge \,
\Gamma_{ {ab}} \, \psi_R \,
+ \, \overline{\psi}_R \, \wedge \,\Gamma_{ {ab}} \, \psi_L
\right)\, \wedge \, V^{ {a}} \, \wedge \,
V^{ {b}}\nonumber\\
&&- \exp\left[ - \, \varphi \right]\, \left(\overline{\psi}_L \,
\wedge \, \Gamma_{ {ab}} \, \rho_R \, + \,
\overline{\psi}_R \, \wedge \,\Gamma_{ {ab}} \, \rho_L
\right)\, \wedge \, V^{ {a}} \, \wedge \,
V^{ {b}}\label{G4Bianchi} \\
0 &=& \mathcal{D}^2\, \chi_{L/R} \, + \, \ft 14 \, R^{ {ab}}
\, \wedge \, \Gamma_{ {ab}} \, \chi_{L/R}\label{dilatinoBianchi}\\
\end{eqnarray}
As it is the case  for all supergravities and for all FDA.s the above
Bianchi identities admit a unique rheonomic solution up to field
redefinitions. The rheonomic solution of the Bianchis implies also the field equations of the theory
given as a set of constraints
to be satisfied by the space-time curvature components. The choice of a frame
 is performed
by imposing an additional  condition which fixes the field
redefinitions. In particular we define the string frame  by  requiring both the
vanishing of the torsion
\begin{equation}
  T^{ {a}} \, = \, 0
\label{zerotorsion}
\end{equation}
and the vanishing of all of the  fermionic sectors of the $3$-form
curvature $\mathbf{H}^{[3]}$. This amounts to setting:
\begin{equation}
  \mathbf{H}^{[3]} \, = \, \mathcal{H}_{abc} \, V^{a} \, \wedge \,
  V^{ {b}} \, \wedge \, V^{ {c}}
\label{stringframedefi}
\end{equation}
One can indeed verify that the fulfillment of the  above
conditions requires a Weyl rescaling of the fields which yields
the usual prefactor $e^{-2\,\varphi}$ in front of the NS-NS and
the fermionic sector of the action. The relevance of the
frame-fixing choice (\ref{stringframedefi}) will be illustrated in
section \ref{GskappaSection} where we discuss the Green-Schwarz
superstring action and $\kappa$-symmetry.
\subsection{Rheonomic parametrizations of the type IIA curvatures in
the string frame}
\label{stringframerheo}
In order to present our result in the form most suitable to our
later purposes, namely the discussion of the BRST chiral algebra
which leads to the construction of the pure spinor superstring action, it is convenient to introduce
a set of tensors, which involve both the supercovariant field
strengths $\mathcal{G}_{ab},\mathcal{G}_{abcd}$ of the Ramond-Ramond $p$-forms and also bilinear currents in the
dilatino field $\chi_{L/R}$. The needed tensors are those listed below:
\begin{eqnarray}
{\mathcal{M}}_{ {ab}} & = & \Big( \ft 18 \, \exp[
\varphi] \, \mathcal{G}_{ {ab}} \, + \, \ft {9}{64} \,
\overline{\chi}_R \, \Gamma_{ {ab}} \, \chi_L \Big)
\nonumber\\
{\mathcal{M}}_{ {abcd}} & = &  - \, \ft 1{16} \, \exp[
\varphi] \, \mathcal{G}_{ {abcd}}
 - \, \ft{3}{256}  \,  \overline{\chi}_L \,
\Gamma_{ {abcd}} \, \chi_R \nonumber\\
\mathcal{N}_0 \, & = &  \ft 34 \, \overline{\chi}_L\, \chi_R \nonumber\\
\mathcal{N}_{ {ab}}&=&
\ft 14  \, \exp[ \varphi] \, \mathcal{G}_{ {ab}} \, + \, \ft{9}{32} \,
\overline{\chi}_R \, \Gamma_{ {ab}}
\, \chi_L=2\,{\mathcal{M}}_{ {ab}} \nonumber\\
\mathcal{N}_{ {abcd}} \, &= &  \ft 1{24} \, \exp[ \varphi] \,
\mathcal{G}_{ {abcd}}
+ \, \ft{1}{128} \, \overline{\chi}_R \,\Gamma_{ {abcd}}
\, \chi_L = - \ft 23 {\mathcal{M}}_{ {abcd}}
\label{Mntensors}
\end{eqnarray}
The above tensors are conveniently assembled into the following
spinor matrices
\begin{eqnarray}
\mathcal{Z} & = & \mathcal{N}_{ {ab}}
\Gamma^{ {ab}}\, + \, 3 \, \mathcal{N}_{ {abcd}}
\,
\Gamma^{ {abcd}}=2\,i\,\mathcal{M}_+\label{Zmatra}\\
\mathcal{M}_\pm &=&{\rm i} \, \left(\mp
\mathcal{M}_{ {ab}} \,
  \Gamma^{ {ab}} \, + \,  \mathcal{M}_{ {abcd}} \,
  \Gamma^{ {abcd}}\right)
\label{Mmatrapm}\\
\mathcal{N}^{(even)}_{\pm} & = & \mp \,\mathcal{N}_0 \, \mathbf{1}
\, + \, \mathcal{N}_{ {ab}} \, \Gamma^{ {ab}} \,
\mp \, \mathcal{N}_{ {abcd}} \, \Gamma^{ {abcd}}
\label{pongo}\\
 \mathcal{N}^{(odd)}_{\pm} & = &\pm \ft
i3\,f_{ {a}}\,\Gamma^{ {a}}\pm\ft{1}{64}\,\overline{\chi}_{R/L}
\, \Gamma_{ {abc}}\,
\chi_{R/L}\,\Gamma^{ {abc}}-\ft i{12}\,\mathcal{H}_{ {abc}}\,\Gamma^{ {abc}}\label{pongo2}\\
\mathcal{L}^{(odd)}_{a\,\pm}&=&
\mathcal{M}_{\mp}\,\Gamma_{ {a}}\,\,;\,\,\,\mathcal{L}^{(even)}_{a\,\pm}=\mp\ft
38\,\mathcal{H}_{ {abc}}\,\Gamma^{ {bc}}
\end{eqnarray}
\par In terms of these objects the rheonomic parametrizations of
the curvatures, solving the Bianchi identities can be written as
follows:
\paragraph{Bosonic curvatures}
\begin{eqnarray}
T^{ {a}} & = & 0 \label{nullatorsioSF}\\
R^{ {ab}} & = & R^{ {ab}}{}_{ {mn}} \,
V^{ {m}} \,\wedge \, V^{ {n}}\, + \,
\overline{\psi}_R\,{\Theta}^{ {ab}}_{ {m}|L}\,
\wedge \, V^{ {m}}\, + \,\overline{\psi}_L\,
{\Theta}^{ {ab}}_{ {m}|R} \, \wedge \, V^{ {m}}\, \nonumber\\
&& + \,{\rm i} \, \ft 34 \, \left( \overline{\psi}_L \, \wedge \,
\Gamma_{ {c}} \, \psi_L \, - \, \overline{\psi}_R \, \wedge
\, \Gamma_{ {c}} \, \psi_R \right)
\, \mathcal{H}^{ {abc}}\nonumber\\
&& \, + \, \overline{\psi}_L \, \wedge \, \Gamma^{[ {a}} \, \mathcal{Z} \,
\Gamma^{ {b}]} \, \psi_R
\label{rheoRiemannSF}\\
\mathbf{H}^{[3]} & = & \mathcal{H}_{ {abc}} V^{ {a}} \, \wedge \, V^{ {b}} \, \wedge \,
V^{ {c}} \label{rheoHSF}\\
\mathbf{G}^{[2]} & = & \mathcal{G}_{ {ab}} V^{ {a}} \, \wedge \, V^{ {b}} \,
\, + \, {\rm i} \, \ft 32 \exp\left[  - \, \varphi \right] \, \left(\overline{\chi}_L  \,
\Gamma_{ {a}} \, \psi_L \,
+\, \overline{\chi}_R \,\Gamma_{ {a}} \, \psi_R
\right)\, \wedge \, V^{ {a}} \label{rheoFSF}\\
\mathbf{f}^{[1]} & = & f_{ {a}} V^{ {a}}  \, + \, \ft 32 \, \left(\overline{\chi}_R  \,
 \psi_L \, - \, \overline{\chi}_L \, \psi_R \right)\label{rheodilatonFSF}\\
 \mathbf{G}^{[4]} & = & \mathcal{G}_{ {abcd}} V^{ {a}} \, \wedge \, V^{ {b}} \, \wedge \,
V^{ {c}} \, \wedge \,
V^{ {d}}\label{rheoGSF} \nonumber\\
&&\, - \, {\rm i} \, \ft 12 \,
\exp[-\varphi] \, \left(\overline{\chi}_L \, \Gamma_{ {abc}} \, \psi_L \, - \,
\overline{\chi}_R \, \Gamma_{ {abc}} \, \psi_R \right) \,
\wedge \, V^{ {a}} \, \wedge \, V^{ {b}} \, \wedge \, V^{ {c}}
\end{eqnarray}
\paragraph{Fermionic curvatures}
\begin{eqnarray}
\rho_{L/R} & = &\rho^{L/R}_{ {ab}} \, V^{ {a}}
\, \wedge \, V^{ {b}} \,
+\mathcal{L}^{(even)}_{a\,\pm}\,\psi_{L/R}\wedge V^a+\mathcal{L}^{(odd)}_{a\,\mp}\,\psi_{R/L}\wedge V^a
\, + \,  \rho_{L/R}^{(0,2)}
 \label{rhoparaSF}\\
\nabla\, \chi_{L/R} & = & \mathcal{D}_{ {a}} \,
\chi_{L/R} \, V^{ {a}}
+\mathcal{N}^{(even)}_{\pm}\,\psi_{L/R}+\mathcal{N}^{(odd)}_{\mp}\,\psi_{R/L}
\label{dechiparaSF}
\end{eqnarray}
Note that the components of the generalized curvatures along the
bosonic vielbeins do not coincide with their spacetime components,
but rather with their supercovariant extension. Indeed expanding
for example the four-form along the spacetime differentials one
finds that \begin{eqnarray} \widetilde G_{\mu\nu\rho\sigma}
&\equiv&\mathcal{G}_{ {abcd}} V^{ {a}}_{\mu} \,
\wedge \, V^{ {b}}_{\nu} \, \wedge \,
V^{ {c}}_{\rho} \, \wedge \, V^{ {d}}_{\sigma} =
\partial_{[\mu}C_{\nu\rho\sigma]}^{[4]} + B_{[\mu\nu}^{[2]}\,\partial_\rho
C^{[1]}_{\sigma]}-\nonumber\\&&-\frac{1}{2}\,e^{-\varphi}\,\left(\overline{\psi}_{L[\mu}\,\Gamma_{\nu\rho}\,\psi_{R\sigma]}+\overline{\psi}_{R[\mu}\,\Gamma_{\nu\rho}\,\psi_{L\sigma]}\right)
+ \, {\rm i} \, \ft 12 \, \exp[-\varphi] \,
\left(\overline{\chi}_L \, \Gamma_{[\mu\nu\rho} \, \psi_{L\sigma]}
\, - \, \overline{\chi}_R \, \Gamma_{[\mu\nu\rho} \,
\psi_{R\sigma]} \right)\nonumber\end{eqnarray} where $\widetilde
G$ is the supercovariant field strength.\par In the
parametrization (\ref{rheoRiemannSF}) of the Riemann tensor we
have used the following definition:
\begin{eqnarray}
\Theta_{ {ab|c}L/R} &=& -i \Big( \Gamma_{ {a}}
\rho_{bc R/L} + \Gamma_{ {b}} \rho_{ca R/L} -
\Gamma_{ {c}} \rho_{ab R/L} \Big)
\end{eqnarray}
Finally by $\rho^{(0,2)}_{L/R}$ we have denoted the fermion-fermion part of
the gravitino curvature whose explicit expression can be written in two
different forms, equivalent by Fierz rearrangement:
\begin{eqnarray}
\rho_{L/R}^{(0,2)}&=& \, \pm \, \ft{21}{32} \, \Gamma_{ {a}} \, \chi_{R/L} \, {\bar \psi}_{L/R} \, \wedge \,
\Gamma^{ {a}} \, \psi_{L/R} \nonumber\\
 && \mp \, \ft{1}{2560} \, \Gamma_{ {a_1a_2a_3a_4a_5}} \, \chi_{R/L} \,  \left (
 \overline{\psi}_{L/R} \,
 \Gamma^{ {a_1a_2a_3a_4a_5}} \, \psi_{L/R} \right )
 \label{rhoparaSF2}\\
 && \mbox{or} \nonumber\\
\rho_{L/R}^{(0,2)}&=& \, \pm \, \ft{3}{8} \, {\rm i}\, \psi_{L/R} \, \wedge \, {\bar \chi}_{R/L} \, \, \psi_{L/R}
  \, \pm \, \ft{3}{16} \, {\rm i}\, \Gamma_{ {ab}} \,\psi_{L/R} \, \wedge \, {\bar \chi}_{R/L} \, \,
  \Gamma^{ {ab}} \, \psi_{L/R}
 \label{rhoparaSF3}
\end{eqnarray}
\subsection{Comments on the curvature structure in the string frame}
The  rheonomic parametrizations presented in the previous section
have some distinctive features which are deprived of any relevance
in a supergravity context while they turn out to be crucial for
the successful construction of a BRST invariant pure spinor
superstring $\sigma$-model. Let us point these features out:
\begin{enumerate}
  \item The rheonomic parametrization of the Neveu-Schwarz curvature
  $\mathbf{H}^{[3]}$ is purely inner, namely there are no dilatino
  terms on the right hand side. As we anticipated this is the very
  definition of the string frame and it is important in order to
  write a $\kappa$-symmetric   Green-Schwarz superstring
  action.
  \item The $(1,1)$ sector of the gravitino curvature
  $\rho^{(1,1)}_{L/R}$ is divided in two parts, one of the same
  chirality, which involves only Neveu-Schwarz field strengths and
  one of the opposite chirality which involves Ramond-Ramond
  field strengths instead:
\begin{equation}
  \rho^{(1,1)}_{L/R} \, = \, \mp \, \underbrace{\ft 38 \, \mathcal{H}_{ {abc}} \, \Gamma^{ {ab}} \,
\psi_{L/R}\, \wedge \, V^{ {c}}}_{\mbox{NS same chirality}}
\, +   \,   \underbrace{\mathcal{M}_\pm \,
 \Gamma_{ {a}} \, \psi_{R/L}
 \, \wedge \,
 V^{ {a}} \,}_{\mbox{RR opposite chirality}}
\label{paffo}
\end{equation}
From a supergravity viewpoint we simply expect a linear combination of
gamma matrices with coefficients given by the bosonic field strengths
and the specific form of such a linear combination has no particular
relevance. On the other hand,  for the
construction of a  pure spinor  BRST invariant action, the particular structure
of $\rho^{(1,1)}_{L/R}$ in the mixed chirality sector, which singles out a matrix $\mathcal{M}_\pm
 \,$ with no vector indices, is just essential. Indeed, as we are going
 to see, the matrix $\mathcal{M}_+$ is just what can be used to
 introduce into the BRST Lagrangian a term of the form:
 $$ \overline{\mathbf{d}}_+ \, \mathcal{M}_+ \, \mathbf{d}_- \, e^+ \, \wedge
 e^-$$
 the fields $\mathbf{d}_\pm$ being the Lagrange multipliers of the
 BRST complex. Such a term  is the vertex operator
 of the Ramond-Ramond fields and it is an important part of
 Berkovits'
 construction. It would not  be allowed if the Lorentz
 structures appearing in $\rho^{(1,1)}_{L/R}$ were different. It
 is remarkable that such a specific Lorentz structure, essential for the pure
 spinor part of the superstring action, appears precisely in the
 string frame, in which the Green-Schwarz part
 of the same superstring action is naturally formulated. For instance in the Einstein frame the
 Lorentz structures appearing in $\rho^{(1,1)}_{L/R}$ are different.
 \item The $\rho^{(0,2)}_{L/R}$ part of the gravitino curvature is
 such that, also in the presence of general backgrounds, with non trivial
 dilatino fields, the contribution to $\rho^{(0,2)}_{L}$ is
 only from bilinears in $\psi_L$ and that to $\rho^{(0,2)}_{R}$
 is only from bilinears in $\psi_R$. This feature is apparent in both the expressions of  $\rho^{(0,2)}_{L/R}$
 given in (\ref{rhoparaSF3}) and will turn out to be crucial in
 proving the BRST invariance of the Berkovits action since it implies
 that the anticommutator of the left-handed BRST operator with the
 right handed one vanishes on the gravitino field. It is once again
 remarkable that this third essential feature of the rheonomic
 parametrizations occurs in the same frame as the other two.  Indeed
 the mentioned structure of $\rho^{(0,2)}_{L/R}$ is not true in the
 Einstein frame.
\end{enumerate}
The above discussion has been  anticipated in order to emphasize
that the subsequent construction of a Berkovits-like pure spinor
superstring action is just founded on the existence of a
supergravity string frame  where the rheonomic parametrizations
display the  three
  features mentioned above.
In solving the Bianchi identities it is by no means  obvious  a
priori that these  features should simultaneously appear. Yet they
do and this gives rise to the Berkovits sigma model.
\subsection{Field equations of type IIA supergravity in the string frame}
\label{equefilde} As usual the rheonomic parametrizations of the
supercurvatures imply, via Bianchi identities a certain number of
constraints on the inner components of the same curvatures which
can be recognized as the field equations of type IIA supergravity.
We derived the  bosonic part of these field equations in two
steps: First we performed the Einstein frame dimensional reduction
on a circle of the field equations of $\mathrm{D=11}$
supergravity. Then we applied the Weyl transformation which
relates the Einstein frame to the string frame:
\begin{equation}
  V_{(E)}^{ {a}} \, = \, V_{(S)}^{ {a}} e^{- \varphi/4}
\label{wile}
\end{equation}
Obviously we could have obtained the same result directly from the
Bianchi identities in the string frame, yet this would have been much
more laborious.
\par The result is the following one. We have an Einstein equation of
the following form:
\begin{eqnarray}
\mbox{\emph{$\mathcal{R}$}}_{ {ab}} & = & \hat T_{ {ab}}\left( f\right) \,  +
\, \hat T_{ {ab}}\left( \mathcal{G}_2\right) \,+ \, \hat T_{ {ab}}\left( \mathcal{H}\right)\, + \, \hat T_{ {ab}}
\left( \mathcal{G}_4\right) \label{Einsteinus}
\end{eqnarray}
where the stress-energy tensor on the right hand side are defined as
\begin{eqnarray}
\hat T_{ {ab}}\left( f\right) & = &\, - \, \mathcal{D}_{ {a}} \, \mathcal{D}_{ {b}}\varphi \,
+ \, \ft 89 \, \mathcal{D}_{ {a}} \, \varphi \, \mathcal{D}_{ {b}} \, \varphi
\, - \, \eta_{ {ab}} \left( \ft 16 \Box \, \varphi \,
+ \, \ft 59 \, \mathcal{D}^{ {m}} \, \varphi \, \mathcal{D}_{ {m}} \, \varphi
\right)\label{dialtostress}\\
\hat T_{ {ab}}\left( \mathcal{G}_2\right) & = &   \exp\left[2 \, \varphi
\right] \, \mathcal{G}_{ {ax}} \, \mathcal{G}_{ {by}} \,
\eta^{ {ab}} \label{Fstresso}\\
\hat T_{ {ab}}\left( \mathcal{H}\right)\, & = & \, - \,\exp\left[\ft 13 \, \varphi
\right] \, \left( \ft 98 \, \mathcal{H}_{ {axy}} \, \mathcal{H}_{ {bwt}} \,
\eta^{ {xw}} \, \eta^{ {yt}}
\, - \, \ft 18 \, \eta_{ {ab}} \, \mathcal{H}_{ {xyz}} \,
\mathcal{H}^{ {xyz}}\right) \label{Hstresso}\\
\hat T_{ {ab}}\left( \mathcal{G}_4\right) & = & \exp\left[2 \, \varphi
\right] \,\left( 6 \,\mathcal{G}_{ {ax_1x_2x_3}} \,
\mathcal{G}_{ {by_1y_2y_3}} \,
 \eta^{ {x_1y_1}} \,
\eta^{ {x_2y_2}}\, \eta^{ {x_3y_3}}\, - \, \ft 12 \,
\eta_{ {ab}} \, \mathcal{G}_{ {x_1\dots x_4}} \,
\mathcal{G}^{ {x_1\dots x_4}}\right)
\end{eqnarray}
Next we have the equations for the dilaton and the Ramond $1$-form:
\begin{eqnarray}
0 & = &\Box \, \varphi \, - \, 2 \,  f_{ {a}} \, f^{ {a}} \, + \, \ft 32 \,
\exp \left[2 \, \varphi \right]
\, \mathcal{G}^{ {x_1x_2}} \, \mathcal{G}_{ {x_1x_2}}\nonumber\\
& & + \, \ft 32 \,
\exp \left[2 \, \varphi \right]
\,
 \mathcal{G}^{ {x_1x_2x_3x_4}} \, \mathcal{G}_{ {x_1x_2x_3x_4}} \,
  + \, \ft 34 \,
\exp \left[\ft 43\, \varphi \right]
\,\mathcal{H}^{ {x_1x_2x_3}} \, \mathcal{H}_{ {x_1x_2x_3}}
\label{Rreq01}\\
0 & = & \mathcal{D}_{ {m}} \,\mathcal{G}^{ {ma}}\,
 - \, \ft 53 \,f^{ {m}} \, \mathcal{G}_{ {ma}}
 \, + \, 3 \, \mathcal{G}^{ {ax_1x_2 x_3}} \, \mathcal{H}_{ {x_1 x_2 x_3}}
 \label{G2equation}
\end{eqnarray}
and the equations for the NS $2$-form and for the RR $3$-form:
\begin{eqnarray}
0 & = & \mathcal{D}_{ {m}} \,\mathcal{H}^{ {mab}}
\, - \, \ft 23 \, f^{ {m}} \,
\mathcal{H}_{ {mab}}\nonumber\\
&&
 \, - \,  \exp\left[  \ft 43 \, \varphi \right]\, \left( 4 \,
\, \mathcal{G}^{ {x_1x_2 ab}} \, \mathcal{G}_{ {x_1 x_2}} \, - \, \ft {1}{24} \,
\epsilon^{ {abx_1 \dots x_8}} \,
\mathcal{G}_{ {x_1x_2x_3x_4}} \,\mathcal{G}_{ {x_5x_6x_7x_8}}\right)
\label{H3equa}\\
0 & = &
\mathcal{D}_{ {m}}\,\mathcal{G}^{ {ma_1 a_2a_3}} \,
+ \, \ft 13 \, f_{m} \, \mathcal{G}^{ {ma_1
a_2a_3}}\nonumber\\
&& + \, \exp\left[  \ft 23 \,
\varphi \right]\, \left( \ft 32 \,
\mathcal{G}^{m[a_1} \, H^{a_2a_3]n} \, \eta_{ {mn}} \,
  \, + \,\ft {1}{48} \, \epsilon^{ {a_1a_2a_3x_1 \dots x_7}} \mathcal{G}_{ {x_1x_2x_3x_4}}
 \, H_{ {x_5x_6x_7}}\right)
\label{23formeque}
\end{eqnarray}
Any solution of these bosonic set of equations can be uniquely
extended to a full superspace solution involving $32$ theta variables
by means of the rheonomic conditions. The implementation of such a
fermionic integration is the \textit{supergauge completion}.
\par
In this way we have completed the discussion of type IIA supergravity
in the string frame. Let us now turn to superstrings.

\newpage
\section{The Green-Schwarz action and $\kappa$-symmetry}
\label{GskappaSection} As we already mentioned in the introduction
the Green-Schwarz $\kappa$-symmetric action of type II
superstrings has exactly the same form (in the string frame of
background supergravity fields) for the IIA and  IIB case. It is
just the form of the $\kappa$-symmetry transformation against
which it is invariant that is slightly different in the two cases.
Actually also these transformations are essentially the same up to
the obvious replacement of $\psi_{L/R}$ gravitinos with their
chiral $\psi_{1,2}$ analogues and similarly for the parameters.
\par
The reason for this equality of the type IIA and type IIB actions
 is due to the following  peculiarity which characterizes both  the NS and  the GS superstring
formalism: they introduce into the action just one half of the
(super)-forms describing (super)-space geometry. All the well
known difficulties connected with the description of RR emission
vertices and with the string quantization in non-trivial RR
backgrounds are connected to this blindness of the formalism which
ignores half of the geometry. The new BRST formulation of
superstring actions based on pure-spinor superghosts is the only,
so far discovered, way-out of this contradiction. Indeed in the
pure spinor approach the ghost-antighost sector appears to provide
the missing fields which couple to the other face of the moon,
namely the fermionic forms $\psi_{L/R}$ and the RR superforms.
Hence the BRST invariant actions of type IIA and type IIB theory
will be different although similar just as the $\kappa$-symmetry
transformations are slightly different in the two cases. The BRST
form of the action is just an extension of the Green-Schwarz
action which is identical in the two cases. This is the world
sheet counterpart of what happens for the bulk supergravity
action. Also there restricting the Lagrangian to the Neveu-Schwarz
sector we obtain  identical sub-Lagrangians while it is the
extension by means of the fermionic and Ramond Ramond fields that
is different in the two cases A and B.
\par
In this section we construct the Green-Schwarz action of type II
superstrings moving in a generic supergravity background and we
 consider its invariance against
$\kappa$-symmetry in the case of type IIA superstrings.
\subsection{The general form of the GS action}
Employing, as it is required by the rheonomic construction of the
Lagrangian, the first order formalism \cite{Fre:2002kc}, we write
the Green-Schwarz action as the sum of two addenda, the kinetic
and the Wess-Zumino contributions:
\begin{equation}
  \mathcal{A}_{GS} \, = \, \mathcal{A}_{kin} \, + \,
  A_{WZ}
\label{AGS2a}
\end{equation}
where:
\begin{eqnarray}
  \mathcal{A}_{kin} & = & \int \, \left( \Pi^{ {a}}_+ \,
  V^{ {b}} \, \eta_{ {ab}} \, \wedge \, e^+ \, - \,
  \Pi^{ {a}}_- \,
  V^{ {b}} \, \eta_{ {ab}} \, \wedge \, e^-
  \,\right.\nonumber\\
  && \left. + \, \ft 12 \Pi^{ {a}}_i\, \Pi^{ {b}}_j
  \, \eta^{ij}\, \eta_{ {ab}} \, e^+ \, \wedge \, e^- \,
 \right )\label{2akinact}\\
A_{WZ} & = & \ft 12 \, q \, \int \, \mathbf{B}^{[2]}
\label{2aWZact}
\end{eqnarray}
In the above two formulae,  $e^\pm \, = \, e^0 \, \pm \, e^1$ denote the zweibein of the
string world-sheet in light-cone basis for the $2d$ Lorentz indices,
namely $\eta_{ij} \, = \, \left(\begin{array}{cc}
  0 & 1 \\
  1 & 0
\end{array} \right) $, while by
$\Pi^{ {a}}_\pm$ we have denoted the usual $0$-form
auxiliary field whose equation identifies it with the projection
of the target vielbein $V^{ {a}}$ onto the world volume
zweibein $e^\pm$. The coefficient $q$, denoting the string charge,
is fixed in such a way as to obtain a completely $\kappa$
supersymmetric action in any background.  As we already stressed
there is no dilaton prefactor in the above action since the FDA
gauge forms (in particular the vielbein) and curvatures were
already transformed to the string frame.
\par
First of all let us check the relative coefficients in the kinetic
action (\ref{2akinact}) by calculating its variation with respect to
the auxiliary field $\Pi^{ {a}}_\pm$. We obtain:
\begin{eqnarray}
0 & = & \frac{\delta\,{\mathcal{A}_{kin}}}{\delta \Pi^{ {a}}_\pm}
 \, = \, \int \,\left(  \pm \, \eta_{ {ab}} \, V^{ {b}} \, \wedge \, e^\pm \, + \,
 \eta_{ {ab}} \, \Pi^{ {b}}_\pm
 \, e^+ \, \wedge \, e^- \right)  \nonumber\\
\null & \Downarrow & \null \nonumber\\
V^{ {a}} & = & \Pi_+^{ {a}} \, e^+ \, + \,
\Pi_-^{ {a}} \, e^-
\label{proieziaPi}
\end{eqnarray}
which is the required result for the elimination of the auxiliary
field $\Pi_-^{ {a}}$ and the transition to second order
formalism.
\par
Next let us introduce the following short hand notation:
\begin{equation}
  \mathbf{\Gamma}_\pm \, \equiv \, \Pi^{a}_\pm \, \Gamma_{ {a}}
\label{shnota}
\end{equation}
and let us  check the $\kappa$-symmetric invariance of
the GS action in the A case.
\subsection{$\kappa$-symmetry in the type IIA case}
Relying on the rheonomic parametrizations of the FDA let us
calculate the  variation of the Green-Schwarz action (\ref{AGS2a})
under a target supersymmetry of parameters $\epsilon_{L/R}$. We
obtain:
\begin{eqnarray}
\delta_{susy} \, \mathcal{A}_{kin} & = & \int \, {\rm i} \left[\left
(
\overline{\epsilon}_L \, \mathbf{\Gamma}_+ \, \psi_L \, \, + \,
\overline{\epsilon}_R \, \mathbf{\Gamma}_+ \, \psi_R\right) \,
\wedge \,  e^+  \,\right.\nonumber\\
&& \left.
 - \, \left (
\overline{\epsilon}_L \, \mathbf{\Gamma}_- \, \psi_L \, \, + \,
\overline{\epsilon}_R \, \mathbf{\Gamma}_- \, \psi_R\right) \,
\wedge \,  e^- \, \right]
\label{variokino2a}
\end{eqnarray}
\begin{eqnarray}
\delta_{susy} \, \mathcal{A}_{WZ} & = & - \, q \,
\int \, {\rm i} \left[\left
(
\overline{\epsilon}_L \, \mathbf{\Gamma}_+ \, \psi_L \, \, - \,
\overline{\epsilon}_R \, \mathbf{\Gamma}_+ \, \psi_R\right) \,
\wedge \,  e^+  \,\right. \nonumber\\
&& \left. + \, \left (
\overline{\epsilon}_L \, \mathbf{\Gamma}_- \, \psi_L \, \, - \,
\overline{\epsilon}_R \, \mathbf{\Gamma}_- \, \psi_R\right) \,
\wedge \,  e^- \, \right] \,
\label{susyWzvario}
\end{eqnarray}
Let us now recall that the rules of the $1.5$-order formalism which we use
in all our $p$-brane constructions impose that, after variation, we
should implement the field equations of all the auxiliary fields
whose equation of motion is algebraic and allows for their own
elimination in terms of dynamical fields. In the string action these
latter are the $0$-form fields $\Pi^{ {a}}_i$ and the
$2$-dimensional zweibein $e^{i}$. The field equation of the first is
(\ref{proieziaPi}) while the field equation of the zweibein is
simply:
\begin{equation}
  \eta_{ {ab}} \, \Pi^{ {a}}_i \,
  \Pi^{ {b}}_j \, = \, \eta_{ij}
\label{inducedmetric}
\end{equation}
namely the statement that the world-sheet metric is the pull-back of
the target superspace  metric. Under these conditions one obtains:
\begin{equation}
  \mathcal{L}_{kin}^{(0)} \, = \, - \, e^+ \, \wedge \, e^-
\label{fantavolume}
\end{equation}
where:
\begin{equation}
  \mathcal{L}_{kin}^{(0)} \, \equiv \,\left( \Pi^{ {a}}_+ \,
  V^{ {b}} \, \eta_{ {ab}} \, \wedge \, e^+ \, - \,
  \Pi^{ {a}}_- \,
  V^{ {b}} \, \eta_{ {ab}} \, \wedge \, e^-
  \, + \, \ft 12 \Pi^{ {a}}_i\, \Pi^{ {b}}_j
  \, \eta^{ij}\, \eta_{ {ab}} \, e^+ \, \wedge \, e^- \,
 \right )
\label{Lnote}
\end{equation}
is the $2$-form which corresponds to the kinetic Lagrangian.
\par
Assembling these results we find:
\begin{eqnarray}
\delta_{susy} \, \mathcal{A}_{GS} & = & \int \,{\rm i}\,
\left[
\, \left( 1-q\right) \, \overline{\epsilon_L}\, \mathbf{\Gamma}_+
\psi_L
\, + \, \left(1+q \right) \overline{\epsilon_R}\, \mathbf{\Gamma}_+
\psi_R \right]  \, \wedge \, e^+\nonumber\\
&& - \, \,{\rm i} \,
 \left[ \left( 1+q\right) \, \overline{\epsilon_L}\,
\mathbf{\Gamma}_-
\psi_L
\, + \, \left(1-q \right) \overline{\epsilon_R}\, \mathbf{\Gamma}_-
\psi_R  \right ]\, \wedge \, e^-
\label{2apantarei}
\end{eqnarray}
The above variation vanishes under the following conditions:
\begin{eqnarray}
q & = & 1 \nonumber\\
\overline{\epsilon}_L & = & \overline{\epsilon}_L \, P^+\nonumber\\
\overline{\epsilon}_R & = & \overline{\epsilon}_R \, P^-
\label{ksusysitua}
\end{eqnarray}
where:
\begin{equation}
  P^\pm \, = \, \ft 12 \left( 1 \, \pm \, \mathbf{\Gamma}_{+-} \right)
\label{KsusyProj}
\end{equation}
is the $\kappa$ supersymmetry projector. Indeed we have
$P^\pm\, \mathbf{\Gamma}_\mp \, = \, 0$
and $P^\pm \, \mathbf{\Gamma}_{+-} \, = \, 1$ which are the
necessary and sufficient conditions in order for half of the terms in
eq.(\ref{2apantarei}) to cancel. The other half of them cancel thanks
to the choice of the parameter $q$.
\par
This concludes the derivation of the $\kappa$-symmetric action of
a type IIA superstring moving in the background of any
supergravity solution, namely of any solution of the type IIA
field equations lifted to the whole $(10,32)$-dimensional
superspace by means of rheonomy. The above formulae encode a
complete algorithm to write down the explicit Green-Schwarz
bosonic-fermionic sigma model once the explicit form of the
superforms $V^{ {a}}\, , \, \mathbf{B}^{[2]}\, ,\,
\varphi$ is given. However, since the most general background is
characterized by mutually interacting fermion, NS-NS and  R-R
fields, these latter have to be determined at the same time as the
NS-NS forms
 and the
fermionic gravitino forms $\psi_{L/R}$.
\subsection{Background independence}
It should be stressed that both in the case of the Green-Schwarz
actions or of their descendant Pure Spinor actions the problem of
constructing the sigma model is always split into two conceptually
well separated parts:
\begin{description}
  \item[a] Construction of the action in a generic supergravity FDA
  background;
  \item[b] Super-gauge completion, namely explicit integration of the
  rheonomic conditions in a given bosonic background in order to produce
  the explicit $\theta$-dependence of the superforms appropriate to
  that background.
\end{description}
The solution of point $[a]$ is universal, can be done once for all
and it is the goal of the present paper. Point $[b]$ is obviously
case dependent and can be more or less technically difficult
depending on the structure of the chosen background. Yet it must
be observed that it is a guaranteed step since the fermionic
equations to be integrated are of the first order and integrable
by very construction. The issue is just a matter of elegance and
brevity in writing the solution, which can always be reached,
although in most cases its explicit expression may require a
considerable calculational effort. We stress this fact because
there has been some confusion about this in the literature,
particularly in connection with the pure spinor formulation. The
pure spinor $\sigma$-model has been constructed case by case on
given backgrounds as if the form of the action and the BRST
transformations had to be reinvented each time. This has probably
somehow obscured the general structure and the remarkable economy
of principles underlying this new setup which solves some of the
open questions in superstring quantization.

\section{The Pure Spinor action and BRST symmetry}
As advocated at the end of the previous section the alternative to
$\kappa$-symmetry is the BRST quantization of the Green-Schwarz
action by means of constrained BRST transformations using pure
spinors superghosts. The procedure consists of the following three
steps:
\begin{description}
  \item[a] Derivation of the constrained BRST algebra in the non-negative
  ghost-number sector from the FDA curvatures and their rheonomic
  parametrizations;
  \item[b] Introduction of antighosts $w^{\pm}$ and Lagrange multipliers
  $\mathbf{d}^\pm$ whose BRST transformation is defined up to a new gauge
  symmetry;
  \item[c] Construction of a \textit{gauge fixing} action
  $\mathcal{A}_{gf}$ to be added to the classical Green-Schwarz
  action $A_{GS}$ such that its variation under BRST cancels
  that of the classical action thanks to the non vanishing BRST
  variation of the Lagrange multipliers amounting to new gauge
  symmetries.
\end{description}
Let us begin with step $[a]$.
\subsection{The constrained BRST algebra from the FDA}
Applying the general procedure we can obtain the explicit form of
the constrained BRST algebra suitable for either the type IIA or the type IIB theory by
performing the ghost-form extension of the Free Differential Algebra
curvature definitions and parametrizations successively setting to
zero the bosonic ghosts. Actually, once the principle has been
clarified we can perform the two steps at once by considering the
purely fermionic extension, namely:
\begin{eqnarray}
\varphi & \mapsto & \varphi \nonumber\\
V^{a} & \mapsto & V^{ {a}} \nonumber\\
\mathbf{B}^{[2]} & \mapsto & \mathbf{B}^{[2]} \nonumber\\
\mathbf{C}^{[1]} & \mapsto & \mathbf{C}^{[1]} \nonumber\\
\mathbf{C}^{[3]} & \mapsto & \mathbf{C}^{[3]} \nonumber\\
\psi_{L/R} & \mapsto & \psi_{L/R} \, + \, \lambda_{L/R}
\label{constghostext2a}
\end{eqnarray}
\par
Each extended curvature definition $\widehat{\mathbf{R}}^{[p]}_{def}$ and each extended curvature
parametrization $\widehat{\mathbf{R}}^{[p]}_{par}$ decomposes into ghost
sectors according to:
\begin{eqnarray}
\widehat{\mathbf{R}}^{[p]}_{def} & = & {\mathbf{R}}^{[p,0]}_{def} \, + \,  {\mathbf{R}}^{[p-1,1]}_{def} \, + \, {\mathbf{R}}^{[p-2,2]}_{def} \,\nonumber\\
\widehat{\mathbf{R}}^{[p]}_{par} & = & {\mathbf{R}}^{[p,0]}_{par} \, + \,  {\mathbf{R}}^{[p-1,1]}_{par} \, + \, {\mathbf{R}}^{[p-2,2]}_{par} \,
\label{curvDefcurvPar}
\end{eqnarray}
where we stop at ghost number $g=2$ since neither in the curvature
definitions nor in the curvature parametrizations there appear higher
than quadratic powers of the $\psi_{L/R}$ forms. Then we have to
impose:
\begin{eqnarray}
{\mathbf{R}}^{[p,0]}_{def} & = & {\mathbf{R}}^{[p,0]}_{par} \nonumber\\
{\mathbf{R}}^{[p-1,1]}_{def} & = & {\mathbf{R}}^{[p-1,1]}_{par}
\nonumber\\
{\mathbf{R}}^{[p-2,2]}_{def} & = & {\mathbf{R}}^{[p-2,2]}_{par}
\label{sectoreque}
\end{eqnarray}
The first of eq.s (\ref{sectoreque}) is simply the rheonomic
parametrization of the classical curvature we started from. The
second equation defines the constrained BRST transformation of all
the physical fields. The last of eq.s (\ref{sectoreque}) defines the
BRST transformation of the ghost fields (the pure spinors) when the
right hand side is non zero ($ {\mathbf{R}}^{[p-2,2]}_{par} \, \ne \,
0$) and the quadratic pure spinor constraints
${\mathbf{R}}^{[p-2,2]}_{def} \, = \, 0$ when the right hand side is
zero ${\mathbf{R}}^{[p-2,2]}_{par}\, = \, 0$.
\par Let us write the result of these straightforward manipulations.
\subsection{Torsionful spin connections}
Before applying the BRST quantization procedures outlined in the
previous subsection it is convenient to rewrite the rheonomic
parametrizations using a differently defined spin connection which
reabsorbs the terms containing the field strength $\mathrm{H}_{abc}$
of the Kalb-Ramond field $B^{[2]}$. This amount to reinterpret
$\mathrm{H}_{abc}$ as the torsion of the ten-dimensional manifold.
\par
To this effect let us consider the structure of the rheonomic
parametrization for the gravitino curvature (\ref{rhoparaSF}). It is
convenient to rewrite the same equations as follows:
\begin{eqnarray}
  \rho^\prime_{L/R}  & \equiv & \mathcal{D}\, \psi_{L/R} \, + \, \, \alpha \,\mathcal{L}^{(even)}_{a\pm}
  \, V^a \, \wedge \, \psi_{L/R}\nonumber \\
  \null &=& \rho^{L/R}_{ab} \, V^a \, \wedge \, V^b \, +(1-\alpha) \, \mathcal{L}^{(even)}_{a\pm}
  \, \psi_{L/R} \wedge \, V^a \, + \, \mathcal{L}^{(odd)}_{a\mp}
  \, \psi_{R/L} \wedge \, V^a \, + \, \rho^{(0,2)}_{L/R} \label{rewritto}
  \end{eqnarray}
  Then let us define a new chiral Lorentz derivative:
  \begin{equation}\label{newderi}
     \nabla^{L/R} \, \equiv \, \mathcal{D} \, + \, \alpha \,\mathcal{L}^{(even)}_{a\pm}
  \, V^a
  \end{equation}
 and let us analyse its properties. Using D=10 spinors we conclude that we have a
 one-parameter family of chiral connections $\nabla^{\alpha}$ of the following form:
 \begin{equation}\label{nablatutto}
    \nabla ^{(\alpha)}\, = \, d \, - \, \ft 14 \left (\omega^{ab} \, \mathbf{1} \, + \,
    \ft 32 \, \alpha \, \mathcal{H}^{abc} \, V_c \, \Gamma_{11} \right ) \, \Gamma_{ab}
 \end{equation}
 Utilizing this notation the rheonomic parametrization of the gravitino is recast into the following
 expression:
 \begin{eqnarray}
   \rho^\prime &\equiv&\nabla ^{(\alpha)}\, \psi= \rho_{ab} \, V^a \, \wedge \, V^b \, - \ft 38\,(1-\alpha)\, \mathcal{H}_{abc} \, \Gamma^{ab} \,
   \Gamma_{11} \, \psi\, \wedge V^c  \\
   \null &\null& \, + \, \mathcal{M} \, \Gamma_a \, \psi \, \wedge \, V^a \, + \, \rho^{(0,2)}
   \label{nablaalphapsi}
 \end{eqnarray}
 The chiral connection cannot be extended to a full-fledged spin connection acting unambiguously also
 on bosonic tensor fields yet on spinors it is perfectly well defined and it can be used to define the BRST operator on fermionic fields. To this effect let us calculate the curvature:
 \begin{eqnarray}
   R^{ab}(\alpha)  & = & \left ( R^{ab} - \, \ft 9{16} \, \alpha^2 \, \mathcal{H}^{amc}
   \, \mathcal{H}^{bmd} \, V_d \,
   \wedge \, V_d\right) \,  \mathbf{1} \nonumber\\
   && + \, \ft 32 \, \alpha \, \left ( \mathcal{D}\mathcal{H} ^{abc} \, V_c \, +
   \mathcal{H} ^{abc} \, \mathcal{D} V_c \right ) \, \Gamma_{11} \label{alfacurva}
 \end{eqnarray}
 Let us consider the $(0,2)$ sector of the curvature $ R^{ab}(\alpha) $. From the rheonomic
 parametrizations we get:
 \begin{eqnarray}
   R^{ab}_{(0,2)}(\alpha) &=& {\rm i} \, \ft 34 \, \left ({\bar \psi}_L \, \wedge \, \Gamma_c \,
   \psi_L \, - \,  {\bar \psi}_R \, \wedge \, \Gamma_c \,
   \psi_R\right ) \, \mathcal{H}^{abc}+\nonumber\\
   \null &\null& {\rm i} \, \ft 34 \,\alpha\,  \left ({\bar \psi}_L \, \wedge \, \Gamma_c \,
   \psi_L \, + \,  {\bar \psi}_R \, \wedge \, \Gamma_c \,
   \psi_R\right ) \, \mathcal{H}^{abc} \, \Gamma_{11} \, + \, {\bar \psi_L} \, \wedge
   \Gamma^{[a} \, \mathcal{Z} \, \Gamma^{b]} \, \psi_R \label{curvettaalpfa}
 \end{eqnarray}
 Hence if we choose $\alpha = -1$ we obtain that on any  chiral spinor $\xi_{L/R}$ the $(0,2)$ sector of the covariant derivative squared behaves as follows:
 \begin{equation}\label{nablasquare}
   \nabla^2_{(0,2)} \xi_{L/R} \, = \, \pm \, {\rm i} \ft 38 \, \left (
   {\bar \psi_{R/L} }\, \Gamma_{c} \, \psi_{R/L}
   H^{abc} \right ) \, \Gamma_{ab} \, \xi_{L/R} \,  - \, \ft 14 \,
   \left ( {\bar \psi_L} \, \wedge
   \Gamma^{[a} \, \mathcal{Z} \, \Gamma^{b]} \, \psi_R \right ) \, \Gamma_{ab} \, \xi_{L/R}
 \end{equation}
 This formula is very important because it shows that upon ghost-extension of the $p$-forms, if
 we define the BRST operator with respect to deformed $\alpha= -1$ connection, then the left
 part of that operator will square to zero on left handed fermions while the right part will square
 to zero on right handed ones.
\subsection{The constrained BRST algebra of type IIA theories}
It is also convenient to split the BRST operator into two chiral
sectors. The BRST operator is written as:
\begin{equation}
  \mathcal{S} \, = \, \mathcal{S}_L \, + \, \mathcal{S}_R
\label{chiralsplit}
\end{equation}
where $\mathcal{S}_{L/R}$ shifts in the direction of $\lambda_{L/R}$.
%
In this way from the $(p-1,1)$ sector we obtain the BRST chiral transformations of
the physical fields:
\begin{eqnarray}
\mathcal{S}_{L/R} \, \mathbf{B}^{[2]}& = & \mp \,2 \,{\rm i} \, \overline{\psi}_{L/R} \, \Gamma_{ {a}} \,
\lambda_{L/R} \, V^{ {a}}  \nonumber\\
\mathcal{S}_{L/R} \, \mathbf{C}^{[1]} & = & \mp \,\exp[-\, \varphi]\,
\overline{\psi}_{R/L} \, \lambda_{L/R} \, + \ft 32 \, {\rm i} \,\exp[-\,
\varphi]\, \overline{\chi}_{L/R} \, \Gamma_{ {a}} \, \lambda_{L/R} \,
V^{ {a}}\nonumber\\
\mathcal{S}_{L/R} \, \mathbf{C}^{[3]} & = & \overline{\psi}_{R/L}
\, \Gamma_{ {ab}} \, \lambda_{L/R} \, V^{ {a}}
\, \wedge \, V^{ {b}}-B^{[2]}\wedge
\mathcal{S}_{L/R}C^{[1]}
\nonumber\\
&& \, \mp  \, {\rm i} \, \ft 12 \, \exp[-\,
\varphi]\, \overline{\chi}_{L/R} \, \Gamma_{ {abc}} \, \lambda_{L/R} \,
V^{ {a}} \, \wedge \, V^{ {b}} \, \wedge \,
V^{ {c}} \nonumber\\
\mathcal{S}_{L/R} \, V^{ {a}} & = & {\rm i} \, \overline{\psi}_{L/R} \,
\Gamma^{ {a}} \, \lambda_{L/R} \nonumber\\
\mathcal{S}_{L/R}\psi_{L/R} & = & - \mathcal{D} \, \lambda_{L/R}
\, \mp \, \ft 38  \, \Gamma^{ {a_1a_2}} \, \lambda_{L/R}
\, V^{ {a_3}}\, \mathcal{H}_{ {a_1a_2a_3}}
\pm\ft{21}{16}\,\Gamma_{ {a}}\chi_{R/L}\,
(\overline{\psi}_{L/R}\,\Gamma^{ {a}}\lambda_{L/R}) \nonumber\\
&& \mp\ft{1}{1280} \, \Gamma_{ {a_1}\dots
 {a_5}}\chi_{R/L}\,
(\overline{\psi}_{L/R}\,\Gamma^{ {a_1}\dots
 {a_5}}\lambda_{L/R})\nonumber\\
\mathcal{S}_{R/L}\psi_{L/R} & = & \mathcal{M}_\pm
 \, \Gamma_{ {b}}
\lambda_{R/L} \, V^{ {b}} \, \label{BRSTtype2aphys}
\end{eqnarray}
while from the sectors $(p-2,2)$ we obtain the transformation of the
superghosts:
\begin{eqnarray}
\mathcal{S}_{L/R}\lambda_{L/R} & =
&\pm\ft{21}{16}\,\Gamma_{ {a}}\chi_{R/L}\,
(\overline{\lambda}_{L/R}\,\Gamma^{ {a}}\lambda_{L/R}) \nonumber\\
&& \mp\ft{1}{1280} \, \Gamma_{ {a_1}\dots
 {a_5}}\chi_{R/L}\,
(\overline{\lambda}_{L/R}\,\Gamma^{ {a_1}\dots
 {a_5}}\lambda_{L/R})
\nonumber\\
\mathcal{S}_{R/L}\lambda_{L/R} & = & 0\label{Sonsupghosts}
\end{eqnarray}
and the following \textit{pure spinor constraints}:
\begin{eqnarray}
0 & = & \overline{\lambda}_L \, \Gamma_{ {a}} \, \lambda_L
\, + \, \overline{\lambda}_R \, \Gamma_{ {a}} \,
\lambda_R  \label{Torsionconstro}\\
0 & = &  \left(  \overline{\lambda}_L \,
\Gamma_{ {a}} \, \lambda_L
\, - \, \overline{\lambda}_R \, \Gamma_{ {a}}
 \, \lambda_R \right) \, \wedge \,
V^{ {a}} \label{B2constro}\\
0 & = &
 \overline{\lambda}_R \,\lambda_L
\label{A1constro}\\
0 & = &
 \overline{\lambda}_R \, \Gamma_{ {ab}} \,\lambda_L \, V^{ {a}}
 \wedge \, V^{ {b}}
\label{C3constro}
\end{eqnarray}
Before discussing the complete structure of the BRST
transformations on the background fields as a consequence of the
extension of the rheonomic parameterizations, we need to clarify
how the constraints (\ref{Torsionconstro})-(\ref{C3constro}) have
to be understood. It is clear that these constraints are too
strong for a 10d target-space vielbein $V^{  a}$ and therefore we
have to project them on the 2d surface by embedding the worldsheet
into the target-space. In particular the vielbeins $V^{ a}$ must
be replaced by the embedding rectangular matrices $\Pi^{ a}_\pm$.
As will be shown in a separate paper \cite{Psconstra}, the set of
constraints given above are equivalent to the constraints given by
\cite{Berkovits:2001ue}. This will be proven by showing that the
solution of the constraints
(\ref{Torsionconstro})-(\ref{C3constro}) gives 22 independent
complex degrees of freedom.\footnote{In \cite{Psconstra} will be
shown that one can obtain a solution of the constraints
(\ref{Torsionconstro})-(\ref{C3constro}) with 22 degrees of
freedom, in a $\mathrm{G}_2$ and in a $\mathrm{SO(8)}$ covariant
basis. Finally, it is proven that the constraints are equivalent
to Berkovits' constraints. As a side result, it is shown that also
the geometrically-deduced constraints for IIA and IIB superstrings
are consistent and equivalent.}~\footnote{The pure spinor
constraints for heterotic strings are derived from superembedding
formalism in \cite{Oda:2001zm,Matone:2002ft}. }

Finally it is also necessary to write down the chiral BRST
transformations of the dilatino field:
\begin{eqnarray}
\mathcal{S}_{L/R} \, \chi_{L/R}& = & \mathcal{N}^{(even)}_\pm \,
\lambda_{L/R}  \nonumber\\
\mathcal{S}_{R/L} \, \chi_{L/R}& = & \, \mathcal{N}^{(odd)}_\mp\,
\lambda_{R/L} \label{Sondilatino}
\end{eqnarray}
Let us give, for the sake of completeness, the formulas defining the
action of the BRST operator on the field strengths, some of which
will be needed in the final section
\begin{eqnarray}
\mathcal{S}_{L/R}\,\mathcal{G}_{ {ab}}&=&e^{-\varphi}\,\left(\pm\overline{\lambda}_{L/R}\,\rho^{R/L}_{ {ab}}-\ft
32\,i\,f_{[ {a}}\,\overline{\chi}_{L/R}\,\Gamma_{ {b}]}\,\lambda_{L/R}+\ft
32\,i\,\mathcal{D}_{[ {a}}\,\overline{\chi}_{L/R}\,\Gamma_{ {b}]}\,\lambda_{L/R}\right.\nonumber\\&&\left.+\ft
32\,i\,\overline{\chi}_{L/R}\,\Gamma_{[ {a}}\,\mathcal{L}^{(even)}_{ {b}]\pm}\lambda_{L/R}+\ft
32\,i\,\overline{\chi}_{R/L}\,\Gamma_{[ {a}}\,\mathcal{L}^{(odd)}_{ {b}]\pm}\lambda_{L/R}\right)\nonumber\\
\mathcal{S}_{L/R}\,\mathcal{G}_{ {abcd}}&=&e^{-\varphi}\,\left(\overline{\lambda}_{L/R}\,\Gamma_{[ {ab}}\rho^{R/L}_{ {cd}]}
\pm\ft
i2\,f_{[ {a}}\,\overline{\chi}_{L/R}\,\Gamma_{ {bcd}]}\,\lambda_{L/R}\mp\ft
i2\,\mathcal{D}_{[ {a}}\,\overline{\chi}_{L/R}\,\Gamma_{ {bcd}]}\,\lambda_{L/R}\right.\nonumber\\&&\left.\mp\ft
i2\,\overline{\chi}_{L/R}\,\Gamma_{[ {abc}}\,\mathcal{L}^{(even)}_{ {d}]\pm}\lambda_{L/R}\pm\ft
i2\,\overline{\chi}_{R/L}\,\Gamma_{[ {abc}}\,\mathcal{L}^{(odd)}_{ {d}]\pm}\lambda_{L/R}-\ft
32\,i\,\mathcal{H}_{[ {abc}}\,
\overline{\chi}_{L/R}\,\Gamma_{ {d}]}\,\lambda_{L/R}\right)\nonumber\\
\mathcal{S}_{L/R}\,\mathcal{H}_{ {abc}}&=& \mp 2
\,i\,\overline{\lambda}_{L/R}\,\Gamma_{[ {a}}\,\rho^{L/R}_{ {bc}]}\nonumber\\
\mathcal{S}_{L/R}\,\mathcal{D}_{ {a}}\chi_{L/R}&=&-\ft
14\,(\overline{\lambda}_{L/R}\,\Theta_{ {cd}, {a}|R/L})\,\Gamma^{ {cd}}\,\chi_{L/R}+
\left[\mathcal{D}_{ {a}}\mathcal{N}^{(even)}_\pm-(\mathcal{N}\,\mathcal{L}_{ {a}})^{(even)}_\pm\right]\,\lambda_{L/R}\nonumber\\
\mathcal{S}_{L/R}\,\mathcal{D}_{ {a}}\chi_{R/L}&=&-\ft
14\,(\overline{\lambda}_{L/R}\,\Theta_{ {cd}, {a}|R/L})\,\Gamma^{ {cd}}\,\chi_{R/L}+
\left[\mathcal{D}_{ {a}}\mathcal{N}^{(odd)}_\pm-(\mathcal{N}\,\mathcal{L}_{ {a}})^{(odd)}_\pm\right]\,\lambda_{L/R}\nonumber\\
\mathcal{S}_{L/R}\,\rho_{ {ab}}^{L/R}&=&\Upsilon^{(even)}_{ {ab}\,\pm}\,\lambda_{L/R}-\ft
14\,
R_{ {cd}, {ab}}\,\Gamma^{ {ab}}\,\lambda_{L/R}+2\,\mathcal{P}_{L/R}[\lambda_{L/R}]\,\rho_{ {ab}}^{L/R}\nonumber\\
\mathcal{S}_{L/R}\,\rho_{ {ab}}^{R/L}&=&\Upsilon^{(odd)}_{ {ab}\,\pm}\,\lambda_{L/R}
\end{eqnarray}
where we have used the following definitions
\begin{eqnarray}
(\mathcal{N}\,\mathcal{L}_{ {a}})^{(odd)}_\pm&\equiv&\mathcal{N}^{(even)}_\mp\,\mathcal{L}_{ {a}\pm}^{(odd)}+
\mathcal{N}^{(odd)}_\pm\,\mathcal{L}_{ {a}\pm}^{(even)}\nonumber\\
(\mathcal{N}\,\mathcal{L}_{ {a}})^{(even)}_\pm&\equiv&\mathcal{N}^{(even)}_\pm\,\mathcal{L}_{ {a}\pm}^{(even)}+
\mathcal{N}^{(odd)}_\mp\,\mathcal{L}_{ {a}\pm}^{(odd)}\nonumber\\
\Upsilon^{(even)}_{ {ab}\,\pm}&=&
\mathcal{D}_{[ {a}}\,\mathcal{L}^{(even)}_{ {b}]\,\pm}+\mathcal{L}^{(even)}_{[ {a}\pm}\,
\mathcal{L}_{ {b}]\pm}^{(even)}+\mathcal{L}^{(odd)}_{[ {a}\mp}\,\mathcal{L}_{ {b}]\pm}^{(odd)}\nonumber\\
\Upsilon^{(odd)}_{ {ab}\,\pm}&=&
\mathcal{D}_{[ {a}}\,\mathcal{L}^{(odd)}_{ {b}]\,\pm}+\mathcal{L}^{(odd)}_{[ {a}\pm}\,
\mathcal{L}_{ {b}]\pm}^{(even)}+\mathcal{L}^{(even)}_{[ {a}\mp}\,\mathcal{L}_{ {b}]\pm}^{(odd)}\nonumber\\
\mathcal{P}_{L/R}[\lambda_{L/R}]&=&\pm\ft
{21}{32}\,\Gamma_{ {a}}\,\chi_{R/L}\,\overline{\lambda}_{L/R}\,\Gamma^{ {a}}\mp\ft
1{2560}\,\Gamma_{ {abcde}}\,\chi_{R/L}\,\overline{\lambda}_{L/R}\,\Gamma^{ {abcde}}\nonumber
\end{eqnarray}
 We have concluded the derivation of the
constrained BRST algebra for
 type IIA superstrings. Let us now go to step [b]
\subsection{The antighosts and the Lagrange multipliers}
The structure of the antighosts and of the Lagrange multipliers is
motivated by the sort of gauging fixing one chooses to implement
on the fermionic symmetries. Let us recall that in flat superspace
the gravitino $1$-form is the exterior derivative of the $\theta$
coordinates:
\begin{equation}
  \psi_{L/R} \, = \, d\theta_{L/R} \quad \mbox{ (flat superspace)}
\label{flatpsi}
\end{equation}
and supersymmetry is nothing else but a translation in
$\theta_{L/R}$:
\begin{equation}
  \theta_{L/R} \, \mapsto \, \theta_{L/R} \, + \, \epsilon_{L/R}
\label{susyshift}
\end{equation}
If we choose, as gauge fixing, the conditions:
\begin{equation}
  \psi_L \, \wedge \, e^+ \, = \, 0 \quad ; \quad \psi_R \, \wedge \,
  e^- \, = \, 0
\label{gaugefixu}
\end{equation}
we obtain that the spinor field $\theta_R$ is holomorphic on the
world sheet while the spinor field $\theta_L$ is antiholomorphic
on it. This is a very good starting point to obtain a
two-dimensional conformal field theory from the pure spinor action
we intend to construct. So, relying on this intuition based on the
case of flat superspace, eq.(\ref{gaugefixu}) is singled out as
our choice. There are no other compelling a-priori reasons to make
such a choice but, once it is made, all the other steps are
essentially determined and lead to an algorithmic derivation of
the action.
\par
Indeed, in order to obtain eq.s (\ref{gaugefixu}) as variational
equations associated with  Lagrange multiplier fields, we decide
that these latter are a pair formed by a left handed
$\mathrm{SO(1,9)}$ spinor $\mathbf{d}_+$ and a right handed
$\mathrm{SO(1,9)}$ spinor $\mathbf{d}_-$ which will finally appear
in the Lagrangian through terms of the following form:
\begin{equation}
  \dots \, + \, \overline{\mathbf{d}}_+ \, \psi_R \, \wedge \, e^+ \, +
  \, \overline{\mathbf{d}}_- \, \psi_L \, \wedge \, e^- \, + \, \dots
\label{Deltaterms}
\end{equation}
This choice determines also the representation assignments of the
antighost fields $w_{\pm}$ which are introduced as those
 chiral spinors of ghost number $g=-1$
 that play the role of predecessors of
the $\mathbf{d}_{\pm}$ fields  through the following relations:
\begin{eqnarray}
\hat{\mathcal{S}}_R \, w_+ & = & \mathbf{d}_+ \nonumber\\
\hat{\mathcal{S}}_L \, w_+& = & 0 \nonumber\\
\hat{\mathcal{S}}_R \, w_- & = &0 \nonumber\\
\hat{\mathcal{S}}_L \, w_-& = & \mathbf{d}_- \label{Sonantigh}
\end{eqnarray}
In the above equations the operators $\widehat{\mathcal{S}}_{L/R}$
denote the chiral parts of the Lorentz covariant BRST operator
obtained from the torsionful connection with $\alpha = -1$:
\begin{equation}\label{splittohatto}
    \widehat{\mathcal{S}} \, \equiv \, {\mathcal{S}}(\alpha=-1) \, = \, \widehat{\mathcal{S}}_L \,
    + \, \widehat{\mathcal{S}}_R
\end{equation}
To motivate this new choice let us observe that on the shell of the
pure spinor constraints the following identity holds true:
\begin{equation}\label{pureshell}
    {\bar \lambda}_L \, \Gamma^a \lambda_L \, = \, - \, {\bar \lambda}_R \, \Gamma^a \lambda_R
    \, \equiv \, X^a
\end{equation}
This identity allows to trade the ordinary BRST operators
$\mathcal{S}_{L/R}$ for the hatted ones which turns out to be very
helpful. Indeed consider the square of the ordinary operator
$\mathcal{S}$. We obtain:
\begin{equation}\label{EsseSquare}
    \mathcal{S}^2 = - \, \ft 14 \, \Gamma_{ab} \, \left ( \ft 32\, {\rm i}\, H^{abc} \, X_c \,
    + \, \ft 12 {\bar \lambda} \Gamma^a \, \mathcal{Z} \, \Gamma^b \, \lambda \, \right )
\end{equation}
On the other hand, as a consequence of (\ref{pureshell}) we also
obtain that the sum $\widehat{\mathcal{S}}_L +
\widehat{\mathcal{S}}_R$ squares to the same expression on an
arbitrary spinor $\phi= \phi_L \, + \, \phi_R$
\begin{equation}\label{uguaglioS}
  \left (\widehat{\mathcal{S}}_L + \widehat{\mathcal{S}}_R\right )^2 \, \phi \, = \,  \mathcal{S}^2
  \, \phi \, = \, - \, \ft 14 \, \left ( \ft 32 \, {\rm i}\, H^{abc} \, X_c \,
    + \, \ft 12 {\bar \lambda} \Gamma^a \, \mathcal{Z} \, \Gamma^b \, \lambda \, \right )\, \Gamma_{ab} \, \phi
\end{equation}
Hence we conclude that on the shell of the pure spinors we have:
\begin{equation}\label{ontheshella}
    {\mathcal{S}}_L + {\mathcal{S}}_R \, = \, \widehat{\mathcal{S}}_L +
    \widehat{\mathcal{S}}_R\label{thesame}
\end{equation}
This  enables us to use either ${\mathcal{S}}_{L/R}$ or
${\widehat{S}}_{L/R}$ according to convenience. For reasons that
will be apparent while constructing the action, in the antighost
sector it is convenient to use the hatted operators.
\par
From eq.(\ref{Sonantigh}) and from eq.(\ref{curvettaalpfa}) one
might conclude that the BRST operators on the fields
$\mathbf{d}_\pm$ necessarily give the following result
\begin{eqnarray}
\widehat{\mathcal{S}}_{R} \, \mathbf{d}_+ & = &  \, {\rm i} \, \ft
38 \,\left ({\bar \lambda}_R \, \Gamma^c \, \lambda_R \right ) \,
\mathcal{H}_{abc} \, \Gamma^{ab} \, w_+
 \nonumber\\
\widehat{\mathcal{S}}_{L} \, \mathbf{d}_- & = & - \, {\rm i} \, \ft
38 \, \left ({\bar \lambda}_R \, \Gamma^c \, \lambda_R \right ) \,
\mathcal{H}_{abc} \, \Gamma^{ab} \, w_-
 \nonumber\\
\widehat{\mathcal{S}}_{L/R}\, \mathbf{d}_\pm &=& - \, \ft 14 \,
{\bar \lambda}_L \, \Gamma^a \, \mathcal{Z} \, \Gamma^b \, \lambda_R
\, \Gamma_{ab} \, w_{\pm} \label{gaugetrasfaDelta}
\end{eqnarray}
but, as already anticipated, this is not the case. Indeed we can
set:
\begin{eqnarray}
\widehat{\mathcal{S}}_{R} \, \mathbf{d}_+ & = &  \xi_+ \, + \, {\rm
i} \, \ft 38 \, \left ({\bar \lambda}_R \, \Gamma^c \, \lambda_R
\right ) \, \mathcal{H}_{abc} \, \Gamma^{ab} \, w_+
 \nonumber\\
\widehat{\mathcal{S}}_{L} \, \mathbf{d}_- & = &\xi_- \, - \, {\rm i}
\, \ft 38 \, \left ({\bar \lambda}_R \, \Gamma^c \, \lambda_R \right
) \, \mathcal{H}_{abc} \, \Gamma^{ab} \, w_-
 \nonumber\\
\widehat{\mathcal{S}}_{L/R}\, \mathbf{d}_\pm &=& - \, \ft 14 \,
{\bar \lambda}_L \, \Gamma^a \, \mathcal{Z} \, \Gamma^b \, \lambda_R
\, \Gamma_{ab} \, w_{\pm} \label{gaugetrasfaDeltaPrime}
\end{eqnarray}
where $\xi_\pm$ encode a new gauge transformation that will be
determined later. Since $\xi_\pm$ are arbitrary parameters we can
redefine them so as to reabsorb the second addend in the first and
the second of equations (\ref{gaugetrasfaDeltaPrime}). So doing we
come to the final form of the BRST transformations on the auxiliary
fields $\mathbf{d}_\pm$:
\begin{eqnarray}
\widehat{\mathcal{S}}_{R} \, \mathbf{d}_+ & = &  \xi_+\nonumber\\
\widehat{\mathcal{S}}_{L} \, \mathbf{d}_- & = &\xi_-\nonumber\\
\widehat{\mathcal{S}}_{L/R}\, \mathbf{d}_\pm &=& - \, \ft 14 \,
{\bar \lambda}_L \, \Gamma^a \, \mathcal{Z} \, \Gamma^b \, \lambda_R
\, \Gamma_{ab} \, w_{\pm} \label{gaugetrasfaDeltaSecond}
\end{eqnarray}
Let us clarify this point. In \cite{Berkovits:2001ue}, the pure
spinor constraints are very simple since they do not interfere with
the background, therefore it is straightforward to derive the gauge
transformations for the conjugate momenta (notice that in
\cite{Berkovits:2001ue} the Hamiltonian formalism has been used). In
our case, the pure spinor constraints
 (\ref{Torsionconstro})-(\ref{C3constro}) involve the vielbein $V^{  a}$ and therefore one
 can wonder what is the interplay with the rest of the action to derive the correct gauge transformations.
 However, one can use the 1.5 formalism
  \cite{VanNieuwenhuizen:1981ae} and consider $V^{  a}$ as  a non-dynamical field, then one derives the gauge transformations
 and, at the end, imposes the equations of motion by replacing $V^{  a}$ by the
 pullbacks $\Pi^{  a}_\pm$. In  \cite{Psconstra}, it is shown that,  by using an adapted basis
 for the pullbacks, the amount of gauge symmetry is the correct one to give 22 degrees of freedom for the conjugate momenta $w_\pm$.

\subsection{The BRST invariant type IIA superstring action}
In \cite{Fre':2006es} we constructed a BRST invariant action for the
M2 brane with pure spinors where we used a certain gauge fixing term.
This construction seems incomplete because it was based on a solution of
the pure spinor constraints which was not complete. There was an idea that the
gauge fixing term could be related to the cohomology class which defines
the FDA but also this idea appears now doubtful. Indeed the M2 brane action we
constructed has no term of the type:
\begin{equation}
  \overline{\mathbf{d}} \, \Gamma_{ {abcd}}\, \cdot \, \mathcal{F}^{ {abcd}} \, \mathbf{d}
\label{DeltaDelta}
\end{equation}
where $\mathbf{d}$ is the Lagrange multiplier field and
$\mathcal{F}_{ {abcd}}$ denotes the $4$-index field
strength. This is a clear indication that the assumptions made
were too restrictive  since a term of the form (\ref{DeltaDelta})
is the vertex of Ramond Ramond fields and it is an essential part
of the Berkovits' superstring Lagrangian. On the other hand this
latter should be related to the M2-brane action by double
dimensional reduction, at least in the case of type IIA theory and
this cannot produce terms of the form (\ref{DeltaDelta}) if they
are missing in higher dimension.
\par
Hence it is  mandatory to repeat the construction of the type IIA
and also of the type IIB pure spinor superstrings from scratch.
Differently from what it was assumed by Berkovits we have shown
that  the constraints are not the same in all cases and, in
particular, they are not the same for type IIB and type IIA
superstrings. Moreover they feel the background and are not given
once for all. In a separate forthcoming publication
\cite{Psconstra} two of us will discuss the solution of the new
formulation of pure spinor constraints streaming from FDA and
rheonomy. Anticipating the result proved in \cite{Psconstra}, we
state that, notwithstanding their different structure the
background dependent constraints derived from the FDA lead to the
same counting of degrees of freedom as in Berkovits' approach both
in the type IIA and type IIB case, namely $22$.  In the case of
type IIA, which is presently under consideration the pure spinor
constraints are given by eq.s
(\ref{Torsionconstro},\ref{B2constro},\ref{A1constro},
\ref{C3constro}). In the presence of these constraints and using
the Lagrange multiplier and antighosts discussed in the previous
section we now construct an addendum
$\mathcal{A}_{gf}^{\mathrm{IIA}}$ to the Green-Schwarz action such
that its BRST variation exactly cancels the BRST variation of the
latter:
\begin{equation}
  \left( \mathcal{S}_L \, + \, \mathcal{S}_R\right) \, \mathcal{A}_{gf}^{\mathrm{IIA}}
  \, = \, - \,
  \left( \mathcal{S}_L \, + \, \mathcal{S}_R\right) \, \mathcal{A}_{GS}
\label{cancellina}
\end{equation}
In order to perform such a construction we begin by writing down
the BRST variation of the Green-Schwarz action. This is
immediately obtained from eq.(\ref{2apantarei}) by replacing the
supersymmetry parameter with the pure spinor superghost and
setting the parameter $q$ to its value $q=1$:
\begin{eqnarray}
\left( \mathcal{S}_L \, + \, \mathcal{S}_R\right)\, \mathcal{A}_{GS}
 & = & \int \, 2\,{\rm i}\, \left[
\,   \overline{\lambda_R}\, \mathbf{\Gamma}_+
\psi_R  \wedge \, e^+
 - \,   \overline{\lambda_L}\, \mathbf{\Gamma}_-
\psi_L \, \wedge \, e^- \right ]
\label{2apantaBRST}
\end{eqnarray}
Next we introduce the following ansatz for the \textit{gauge fixing}
action:
\begin{eqnarray}
\mathcal{A}_{gf}^{\mathrm{IIA}} & = & \mathcal{S}_{R} \, \left(
\overline{w}_+ \, \psi_R \, \wedge \, e^+ \right) \, + \,
\mathcal{S}_{L} \, \left(
\overline{w}_- \, \psi_L \, \wedge \, e^- \right) \nonumber\\
&& \, + \, \mathcal{S}_{R}\, \mathcal{S}_{L} \, \left (
\overline{w}_{+} \, \Omega \, w_- \, e^+ \, \wedge \, e^- \right)
\end{eqnarray}
where   $\Omega$ is a matrix in spinor space constructed by
saturating gamma matrices only with physical curvature components.
The precise form of  $\Omega$ will now be determined by imposing eq.
(\ref{cancellina}). In the following we shall use the hatted BRST
operators in virtue of the property (\ref{thesame}). The fact that
$\Omega$ depends on physical fields only implies that the action of
the operators $\widehat{\mathcal{S}}_L^2$ and
$\widehat{\mathcal{S}}_R^2$ on them are zero modulo Lorentz
transformations and this allows the following formal manipulations:
\begin{eqnarray}
\left( \widehat{\mathcal{S}}_L \, + \,
\widehat{\mathcal{S}}_R\right) \, A_{gf}^{\mathrm{IIA}} & = &
\widehat{\mathcal{S}}_R^2 \, \left( \overline{w}_+ \, \psi_R \,
\wedge \, e^+ \right) \, + \, \widehat{\mathcal{S}}_L^2 \,
\left( \overline{w}_- \, \psi_L \, \wedge \, e^- \right) \nonumber\\
 && \, - \, \widehat{\mathcal{S}}_R \,\, \widehat{\mathcal{S}}_L
\left( \overline{w}_+ \, \psi_R \, \wedge \, e^+ \right) \,  - \,
\widehat{\mathcal{S}}_L\, \widehat{\mathcal{S}}_R \left(
\overline{w}_+ \, \psi_R \, \wedge \, e^+ \right)
\nonumber\\
&& \, + \, \widehat{\mathcal{S}}_L \,\widehat{\mathcal{S}}_R^2 \,
\left ( \overline{w}_{+} \,
 \Omega \, w_- \, e^+ \, \wedge \, e^- \right)\nonumber\\
&& \, - \, \widehat{\mathcal{S}}_R \, \widehat{\mathcal{S}}_L^2
\left ( \overline{w}_{+} \,
 \Omega \, w_- \, e^+ \, \wedge \, e^- \right)
\label{formamanipula}
\end{eqnarray}
Next taking into account that the only field on which
$\widehat{\mathcal{S}}_{L/R}^2$ is non zero is $w_\mp$, from
eq.(\ref{formamanipula}) we obtain:
\begin{eqnarray}
\left( \widehat{\mathcal{S}}_L \, + \,
\widehat{\mathcal{S}}_R\right) \, A_{gf}^{\mathrm{IIA}} & = &
\overline{\xi}_+ \, \psi_R \, \wedge \, e^+ \, + \,
\xi_-\psi_L \, \wedge \, e^-\nonumber\\
&& \, + \, \widehat{\mathcal{S}}_R \, \left[\overline{w}_+ \,
\widehat{\mathcal{S}}_L \,\left( \psi_R \right)
\, \wedge \, e^+ \, - \, \overline{w}_+ \, \Omega \, \xi_- \, e^+ \, \wedge \, e^- \right] \nonumber\\
&&\, + \, \widehat{\mathcal{S}}_L \, \left[\overline{w}_- \,
\widehat{\mathcal{S}}_R \,\left( \psi_L \right) \, \wedge \, e^- \,
+ \, \overline{\xi}_+ \, \Omega \, w_- \, e^+ \, \wedge \, e^-
\right]\,. \label{formamanipula2}
\end{eqnarray}
Combining these results with the BRST variation of the
Green-Schwarz action given in eq.(\ref{2apantaBRST}) conclude that
we have BRST invariance of the complete action, namely:
\begin{equation}
  \left( \widehat{\mathcal{S}}_L \, + \, \widehat{\mathcal{S}}_R\right) \, \left(
  \mathcal{A}_{GS} \, + \,  \mathcal{A}_{gf}^{\mathrm{IIA}}\right) \, = \, 0
\label{canzelloIIa}
\end{equation}
 if the following conditions are satisfied :
\begin{eqnarray}
\overline{\xi}_+ \, \psi_R \, \wedge \, e^+ & = &
 - \, 2 \, {\rm i}  \overline{\lambda}_R
\, \mathbf{\Gamma}_+ \, \psi_R \, \wedge \, e^+ \,,
\label{picco1}\\
\overline{\xi}_- \, \psi_L \, \wedge \, e^- & = &
  \, 2 \, {\rm i}  \overline{\lambda}_L
\, \mathbf{\Gamma}_- \, \psi_L \, \wedge \, e^- \,, \label{picco2}
\end{eqnarray}
and moreover if the arguments of $\widehat{\mathcal{S}}_{R/L}$ in
the last two lines of eq. (\ref{formamanipula2}) vanish separately.
Conditions (\ref{picco1}), (\ref{picco2}) allow to determine the
gauge transformation of the  anti-ghost fields, namely $\xi_\pm$. We
indeed find:
 \begin{eqnarray}
\bar{\xi}_+&=&-2i\,\bar{\lambda}_R\,\mathbf{\Gamma}_+\,\,;\,\,\,\bar{\xi}_-=2i\,\bar{\lambda}_L\mathbf{\Gamma}_-\,,
\end{eqnarray}
or, equivalently,
 \begin{eqnarray}
\xi_+&=&2i\,\mathbf{\Gamma}_+\,\lambda_R\,\,;\,\,\,\xi_-=-2i\,\mathbf{\Gamma}_-\,\lambda_L\,.
\end{eqnarray}
We next require the vanishing of the arguments of
$\widehat{\mathcal{S}}_{R/L}$ in the last two lines of eq.
(\ref{formamanipula2}). This implies
\begin{eqnarray}
0&=&\overline{w}_+ \, \widehat{\mathcal{S}}_L \,\left( \psi_R
\right) \, \wedge \, e^+ -\, \overline{w}_+ \, \Omega \, \xi_- \,
e^+ \wedge \, e^-= \nonumber\\&=&\overline{w}_+ \, \mathcal{M}_- \,
\mathbf{\Gamma}_-\,\lambda_L  \,e^+ \wedge \, e^- -2\,i\,
\overline{w}_+ \, \Omega \, \mathbf{\Gamma}_-\,\lambda_L \, e^+
\wedge \, e^-\,, \label{cucco1}
\end{eqnarray}
where we have used the last of equations (\ref{BRSTtype2aphys}) to
express $\widehat{\mathcal{S}}_{L/R} \,\left( \psi_{R/L} \right)$.
Equation (\ref{cucco1}) is satisfied provided we make the following
identification:
\begin{eqnarray}
\mathcal{M}_- &=&2\,i\,\Omega\,.\label{mom}
\end{eqnarray}
The second condition reads:
\begin{eqnarray}
 0&=&\overline{w}_- \, \widehat{\mathcal{S}}_R \,\left( \psi_L
\right) \, \wedge \, e^-+  \overline{\xi}_+ \, \Omega \, w_- \,
e^+ \, \wedge \, e^- =\nonumber\\&=&\overline{w}_-
\,\mathcal{M}_+\,\mathbf{\Gamma}_+\,\lambda_R\,e^+ \, \wedge \,
e^--2\,i\,\bar{\lambda}_R\,\mathbf{\Gamma}_+\,\Omega\,w_-\,e^+ \,
\wedge \, e^-\,.\label{cucco2}
\end{eqnarray}
Now we may use the following property:
\begin{eqnarray}
\overline{w}_-
\,\mathcal{M}_+\,\mathbf{\Gamma}_+\,\lambda_R&=&w_-^T\,C\,\mathcal{M}_+\,\mathbf{\Gamma}_+\,\lambda_R=\lambda_R^T\,C\,
\mathbf{\Gamma}_+\,C^{-1}\,\mathcal{M}_+^T\,C\,w_-=\bar{\lambda}_R\,\mathbf{\Gamma}_+\,\tilde{\mathcal{M}_+}\,w_-\,,
\end{eqnarray}
where $C$ denotes the charge conjugation matrix, defined by the
property $C^{-1}\,\Gamma_{ {a}}\,C=-\Gamma_{ {a}}^T$, and
$\tilde{\mathcal{M}_\pm}=C^{-1}\,\mathcal{M}_\pm^T\,C$. Equation
(\ref{cucco2}) then implies
\begin{eqnarray}
\tilde{\mathcal{M}_+}&=&2\,i\,\Omega\,,
\end{eqnarray}
which is consistent with (\ref{mom}) since
\begin{eqnarray}
\tilde{\mathcal{M}_\pm} &=&\mathcal{M}_\mp\,.
\end{eqnarray}

\subsection{Explicit form of the Pure Spinor $\sigma$-model action}

Here we  explicitly compute the terms coming form new piece of the
action denoted by ${\cal A}^{\mathrm IIA}_{gf}$ by acting with the
BRST operators $\widehat{\mathcal S}_L$ and $\widehat{\mathcal S}_R$
on the "gauge-fixing" terms
\begin{eqnarray}\label{expA}
\mathcal{A}_{gf}^{\mathrm{IIA}} & = & \widehat{\mathcal{S}}_{R} \,
\left( \overline{w}_+ \, \psi_R \, \wedge \, e^+ \right) \, + \,
\mathcal{S}_{L} \, \left(
\overline{w}_- \, \psi_L \, \wedge \, e^- \right) \nonumber\\
&& \, + \, \widehat{\mathcal{S}}_{R}\, \widehat{\mathcal{S}}_{L} \,
\left ( \overline{w}_{+} \, \Omega \, w_- \, e^+ \, \wedge \, e^-
\right)
\nonumber\\
&=&
\overline{\mathbf d}_+ \, \psi_R \, \wedge \, e^+ +
\overline{\mathbf d}_-  \, \psi_L \, \wedge \, e^-   +  \frac{\rm i}{2}
\overline{\mathbf d}_+  \,  \mathcal{M}_-  \, {\mathbf  d}_- \nonumber \\
&-& \overline{w}_+ \left(\widehat{\mathcal{S}}_{R}\psi_R\right) \,
\wedge \, e^+
- \overline{w}_- \, \left(\widehat{\mathcal{S}}_{L}  \psi_L\right) \, \wedge \, e^-  \nonumber \\
&-& \frac{\rm i}{2} \,  \overline{w}_+  \left(
\widehat{\mathcal{S}}_{R} \mathcal{M}_-  \right) {\mathbf  d}_- +
\frac{\rm i}{2} \, \overline{\mathbf d}_+  \left(
\widehat{\mathcal{S}}_{L} \mathcal{M}_- \right) {w}_- - \frac{\rm
i}{2} \,  \overline{w}_+ \left( \widehat{\mathcal{S}}_{R}
\widehat{\mathcal{S}}_{L} \mathcal{M}_-  \right) {w}_-\nonumber\\
&+&{\rm i} \, \ft 18 \, {\bar w}_+ \mathcal{M}_ -\, \Gamma_{ab} \,
w_-\, {\bar \lambda}_L \Gamma^a \, \mathcal{Z} \, \Gamma^b \,
\lambda_R
\end{eqnarray}
where the action of ${\mathcal{S}}$ on $\psi_{L/R}$ is given in
(\ref{BRSTtype2aphys}), while the action of
$\widehat{\mathcal{S}}_{L/R}$ on the spinor matrices
$\mathcal{M}_\pm$ can be deduced by computing the corresponding BRST
variation of the tensors in (\ref{Mntensors}), which reads as
follows
\begin{eqnarray}\label{smm}
\mathcal{S}_{L/R}\mathcal{M}_-&=&\pm\ft
i8\,\overline{\lambda}_{L/R}\,\rho^{R/L}_{ {ab}}\,\Gamma^{
{ab}}-\ft i{16}\,\overline{\lambda}_{L/R}\,\Gamma_{
{ab}}\rho^{R/L}_{ {cd}}\,\Gamma^{ {abcd}}- \ft
3{16}\,\overline{\lambda}_{L/R}\,\Gamma_{[ {a}}\,\mathcal{D}_{
{b}]}\chi_{L/R}\,\Gamma^{ {ab}}\nonumber\\&&\pm \ft
1{32}\,\overline{\lambda}_{L/R}\,\Gamma_{[ {abc}}\,\mathcal{D}_{
{d}]}\chi_{L/R}\,\Gamma^{ {ab}}+
\overline{\chi}_{R}\,\mathcal{A}^-_{L\,\vert\lambda_{R/L}=0}+\overline{\chi}_{L}\,\mathcal{A}^-_{R\,\vert\lambda_{R/L}=0}
\end{eqnarray}
where we have defined $\mathcal{A}^-_{L/R}$ in the following way
\begin{eqnarray}
\mathcal{A}^-_{L/R}&=&\left(\pm \ft
3{16}\,i\,\lambda_{L/R}\,e^{\varphi}\,G_{ {ab}}+\ft 3{16}\,f_{[
{a}}\,\Gamma_{ {b}]}\,\lambda_{R/L}-\ft 3{16}\,\Gamma_{[
{a}}\mathcal{L}^{(even)}_{ {b}]\mp}\,\lambda_{R/L}-\ft
3{16}\,\Gamma_{[ {a}}\mathcal{L}^{(odd)}_{
{b}]\pm}\,\lambda_{L/R}\right.\nonumber\\&&\left.
\pm\ft{9}{64}\,i\,\Gamma_{ {ab}}
\mathcal{N}^{(even)}_{\pm}\,\lambda_{L/R}
\pm\ft{9}{64}\,i\,\Gamma_{ {ab}}
\mathcal{N}^{(odd)}_{\mp}\,\lambda_{R/L} \right)\otimes \Gamma^{
{ab}}\nonumber\\&&+\left(\mp\ft
3{32}\,i\,\lambda_{L/R}\,e^{\varphi}\,G_{ {abcd}}- \ft
3{32}\,\mathcal{H}_{[ {abc}}\,\Gamma_{ {d}]}\,\lambda_{R/L}\pm \ft
1{32}\, \Gamma_{[ {abc}}\mathcal{L}^{(even)}_{
{d}]\mp}\,\lambda_{R/L}\right.\nonumber\\&&\left.\pm \ft 1{32}\,
\Gamma_{[ {abc}}\mathcal{L}^{(odd)}_{
{d}]\pm}\,\lambda_{L/R}\mp\ft 1{32}\,f_{[ {a}}\,\Gamma_{
{bcd}]}\,\lambda_{R/L}-\ft{3\,i}{256}\,\Gamma_{ {abcd}}
\mathcal{N}^{(even)}_{\pm}\,\lambda_{L/R}\right.\nonumber\\&&\left.-\ft{3\,i}{256}\,\Gamma_{
{abcd}}
\mathcal{N}^{(odd)}_{\mp}\,\lambda_{R/L}\right)\,\otimes\,\Gamma^{
{abcd}}
\end{eqnarray}
The complete expression of
$\mathcal{S}_R\,\mathcal{S}_L\,\mathcal{M}_-$, which can be computed
using the above formulas, is rather involved. Therefore we shall
give it below for $\chi=0$
\begin{eqnarray}\label{spp}
\left[\mathcal{S}_R\,\mathcal{S}_L\,\mathcal{M}_-\right]_{\chi=0}&=&-\ft
i8\,\overline{\lambda}_L\,\left(\Upsilon^{(even)}_{ {ab}-}-\ft
14\,R_{ {cd}, {ab}}\,\Gamma^{ {cd}}\right)\,\lambda_R\,\Gamma^{
{ab}}\nonumber\\&&+\ft i{16}\,\overline{\lambda}_L\,\Gamma_{
{ab}}\,\left(\Upsilon^{(even)}_{ {cd}-}-\ft 14\,R_{ {ef},
{cd}}\,\Gamma^{ {ef}}\right)\,\lambda_R\,\Gamma^{
{abcd}}\nonumber\\&&+\ft 3{16}\,\overline{\lambda}_L\,\Gamma_{
{a}}\, \left(\mathcal{D}_{
{b}}\mathcal{N}^{(odd)}_--(\mathcal{N}\,\mathcal{L}_{
{b}})^{(odd)}_-\right)\,\lambda_R\,\Gamma^{ {ab}}
\nonumber\\&&-\ft 1{32}\,\overline{\lambda}_L\,\Gamma_{ {abc}}\,
\left(\mathcal{D}_{
{d}}\mathcal{N}^{(odd)}_--(\mathcal{N}\,\mathcal{L}_{
{d}})^{(odd)}_-\right)\,\lambda_R\,\Gamma^{ {abcd}}+
\overline{\lambda}_R\tilde{\mathcal{N}^{(even)}_-}\,\mathcal{A}^-_{L\,\vert
\lambda_R=0}\nonumber\\&&+\overline{\lambda}_R\tilde{\mathcal{N}^{(odd)}_-}\,\mathcal{A}^-_{R\,\vert
\lambda_R=0}
\end{eqnarray}
\paragraph{The quartic ghost interactions}
The action we have displayed contains a battery of terms of the form
$w_+ \, w_- \, \lambda_L \, \lambda_R$. They have two origins.
 On one side they are generated by $\overline{w}_+\left[\mathcal{S}_R\,\mathcal{S}_L\,\mathcal{M}_-\right]w_-$, as we have shown above,
on the other side they come from $\overline{w}_+\,\mathcal{M}_-
\,\mathcal{S}_R\,\mathcal{S}_L\, w_-$.
\par
On every particular supergravity background one has to calculate the
contribution of both sources of quartic ghost terms.
\section{Conclusions}

We provided a complete geometrical derivation of the pure spinor
sigma model for type IIA superstrings based on the FDA of the
corresponding supergravity. The FDA formulation of the latter had to
be adapted to this problem by using directly the string frame rather
than the Einstein frame. This require a field redefinition. It
turned out that the solution of the Bianchi identities and the
construction of the supergravity  rheonomic parametrization was much
easier derived directly in the string frame than obtained it from
field redefinitions and dimensional reduction starting from 11d.
From this effort, we gained a very simple rule for the BRST
transformations to be used for the pure spinor formulation. The
latter is obtained in a Lagrangian formalism and the result has the
advantage to relate the superfields appearing in the FDA with those
appearing in the BRST transformation and in the sigma model. That is
important in order to have  a straight path for constructing the
sigma model given in any supergravity background.

\section*{Acknowledgments}
We thank L. Castellani and G. Policastro for very useful discussions.
\appendix
\section{Summary Tables}

\begin{table}[!htb]
  \caption{\textbf{Tensors and Matrices: }{\it Recalling that $\mathcal{G}_{ {ab}}$ and $\mathcal{G}_{ {abcd}}$ denote
  the supercovariant
  field strengths of the Ramond Ramond $1$-form and $3$-form respectively, $\mathcal{H}_{ {abc}}$ the supercovariant field strength
  of the Neveu Schwarz two-form, while $\chi_{L/R}$ denote
  the chiral components of the dilatino spinor field and $\varphi,f_{ {a}}$ denote the dilaton and its supercovariant derivative,
  the table below summarizes
  the precise definition of certain tensors and matrices appearing both in the sigma model action and in the
  BRST transformation rules.}}\label{tbl:tensor}
\begin{center}
$
\begin{array}{|rcl|}
\hline
{\mathcal{M}}_{ {ab}} & = & \Big( \ft 18 \, \exp[
\varphi] \, \mathcal{G}_{ {ab}} \, + \, \ft {9}{64} \,
\overline{\chi}_R \, \Gamma_{ {ab}} \, \chi_L \Big)
\nonumber\\
{\mathcal{M}}_{ {abcd}} & = &  - \, \ft 1{16} \, \exp[
\varphi] \, \mathcal{G}_{ {abcd}}
 - \, \ft{3i}{256}  \,  \overline{\chi}_L \,
\Gamma_{ {abcd}} \, \chi_R \nonumber\\
\mathcal{N}_0 \, & = &  \ft 34 \, \overline{\chi}_L\, \chi_R \nonumber\\
\mathcal{N}_{ {ab}}&=&
\ft 14  \, \exp[ \varphi] \, \mathcal{G}_{ {ab}} \, + \, \ft{9}{32} \,
\overline{\chi}_R \, \Gamma_{ {ab}}
\, \chi_L=2\,{\mathcal{M}}_{ {ab}} \nonumber\\
\mathcal{N}_{ {abcd}} \, &= &  \ft 1{24} \, \exp[ \varphi] \,
\mathcal{G}_{ {abcd}}
+ \, \ft{1}{128} \, \overline{\chi}_R \,\Gamma_{ {abcd}}
\, \chi_L = - \ft 23 {\mathcal{M}}_{ {abcd}}\\
\hline
\mathcal{Z} & = & \mathcal{N}_{ {ab}}
\Gamma^{ {ab}}\, + \, 3 \, \mathcal{N}_{ {abcd}}
\,
\Gamma^{ {abcd}}\\
\mathcal{M}_\pm &=&{\rm i} \, \left(\mp
\mathcal{M}_{ {ab}} \,
  \Gamma^{ {ab}} \, + \,  \mathcal{M}_{ {abcd}} \,
  \Gamma^{ {abcd}}\right)
\\
\mathcal{N}^{(even)}_{\pm} & = & \mp \,\mathcal{N}_0 \, \mathbf{1}
\, + \, \mathcal{N}_{ {ab}} \, \Gamma^{ {ab}} \,
\mp \, \mathcal{N}_{ {abcd}} \, \Gamma^{ {abcd}}
\\
 \mathcal{N}^{(odd)}_{\pm} & = &\pm \ft
i3\,f_{ {a}}\,\Gamma^{ {a}}\pm\ft{1}{64}\,\overline{\chi}_{R/L}
\, \Gamma_{ {abc}}\,
\chi_{R/L}\,\Gamma^{ {abc}}-\ft i{12}\,\mathcal{H}_{ {abc}}\,\Gamma^{ {abc}}\\
\mathcal{L}^{(odd)}_{a\,\pm}&=&
\mathcal{M}_{\mp}\,\Gamma_{ {a}}\,\,;\,\,\,\mathcal{L}^{(even)}_{a\,\pm}=\mp\ft
38\,\mathcal{H}_{ {abc}}\,\Gamma^{ {bc}}\\
\hline
(\mathcal{N}\,\mathcal{L}_{ {a}})^{(odd)}_\pm&\equiv&\mathcal{N}^{(even)}_\mp\,\mathcal{L}_{ {a}\pm}^{(odd)}+
\mathcal{N}^{(odd)}_\pm\,\mathcal{L}_{ {a}\pm}^{(even)}\\
(\mathcal{N}\,\mathcal{L}_{ {a}})^{(even)}_\pm&\equiv&\mathcal{N}^{(even)}_\pm\,\mathcal{L}_{ {a}\pm}^{(even)}+
\mathcal{N}^{(odd)}_\mp\,\mathcal{L}_{ {a}\pm}^{(odd)}\\
\Upsilon^{(even)}_{ {ab}\,\pm}&=&
\mathcal{D}_{[ {a}}\,\mathcal{L}^{(even)}_{ {b}]\,\pm}+\mathcal{L}^{(even)}_{[ {a}\pm}\,
\mathcal{L}_{ {b}]\pm}^{(even)}+\mathcal{L}^{(odd)}_{[ {a}\mp}\,\mathcal{L}_{ {b}]\pm}^{(odd)}\\
\Upsilon^{(odd)}_{ {ab}\,\pm}&=&
\mathcal{D}_{[ {a}}\,\mathcal{L}^{(odd)}_{ {b}]\,\pm}+\mathcal{L}^{(odd)}_{[ {a}\pm}\,
\mathcal{L}_{ {b}]\pm}^{(even)}+\mathcal{L}^{(even)}_{[ {a}\mp}\,\mathcal{L}_{ {b}]\pm}^{(odd)}\\
\mathcal{P}_{L/R}[\lambda_{L/R}]&=&\pm\ft
{21}{32}\,\Gamma_{ {a}}\,\chi_{R/L}\,\overline{\lambda}_{L/R}\,\Gamma^{ {a}}\mp\ft
1{2560}\,\Gamma_{ {abcde}}\,\chi_{R/L}\,\overline{\lambda}_{L/R}\,\Gamma^{ {abcde}}\\
\hline
\end{array}$
\end{center}
\end{table}
\begin{table}[!htb]
  \caption{\textbf{BRST algebra: }{\it In this table we summarize the BRST transformations
  of the fundamental fields. In the first box are displayed the BRST transformations of the
  physical fields encoded in the supergravity forms: the vielbein $V^{a}$, the NS $2$-form $\mathbf{B}^{[2]}$, the
  Ramond Ramond forms $\mathbb{C}^{[1,3]}$ and the gravitino $\psi_{L/R}$. In the second box those of
  the (pure spinor) superghosts $\lambda_{L/R}$ , while the third
  box gives the transformations of the antighosts $w_\pm$ and of the Lagrange multipliers $\mathbf{d}_\pm$  .}}\label{tbl:transfe}
\begin{center}
$
\begin{array}{|rcl|}
\hline
\mathcal{S}_{L/R} \, \mathbf{B}^{[2]}& = & \mp \,2 \,{\rm i} \, \overline{\psi}_{L/R} \, \Gamma_{ {a}} \,
\lambda_{L/R} \, V^{ {a}} \\
\mathcal{S}_{L/R} \, \mathbf{C}^{[1]} & = & \mp \,\exp[-\, \varphi]\,
\overline{\psi}_{R/L} \, \lambda_{L/R} \, + \ft 32 \, {\rm i} \,\exp[-\,
\varphi]\, \overline{\chi}_{L/R} \, \Gamma_{ {a}} \, \lambda_{L/R} \,
V^{ {a}}\\
\mathcal{S}_{L/R} \, \mathbf{C}^{[3]} & = & \overline{\psi}_{R/L}
\, \Gamma_{ {ab}} \, \lambda_{L/R} \, V^{ {a}}
\, \wedge \, V^{ {b}}-B^{[2]}\wedge
\mathcal{S}_{L/R}C^{[1]}
\\
&& \, \mp  \, {\rm i} \, \ft 12 \, \exp[-\,
\varphi]\, \overline{\chi}_{L/R} \, \Gamma_{ {abc}} \, \lambda_{L/R} \,
V^{ {a}} \, \wedge \, V^{ {b}} \, \wedge \,
V^{ {c}} \\
\mathcal{S}_{L/R} \, V^{ {a}} & = & {\rm i} \, \overline{\psi}_{L/R} \,
\Gamma^{ {a}} \, \lambda_{L/R} \\
\mathcal{S}_{L/R}\psi_{L/R} & = & - \mathcal{D} \, \lambda_{L/R}
\, \mp \, \ft 38  \, \Gamma^{ {a_1a_2}} \, \lambda_{L/R}
\, V^{ {a_3}}\, \mathcal{H}_{ {a_1a_2a_3}}
\pm\ft{21}{16}\,\Gamma_{ {a}}\chi_{R/L}\,
(\overline{\psi}_{L/R}\,\Gamma^{ {a}}\lambda_{L/R}) \\
&& \mp\ft{1}{1280} \, \Gamma_{ {a_1}\dots
 {a_5}}\chi_{R/L}\,
(\overline{\psi}_{L/R}\,\Gamma^{ {a_1}\dots
 {a_5}}\lambda_{L/R})\\
\mathcal{S}_{R/L}\psi_{L/R} & = & \mathcal{M}_\pm
 \, \Gamma_{ {b}}
\lambda_{R/L} \, V^{ {b}} \\
\hline
\mathcal{S}_{L/R}\lambda_{L/R} & =
&\pm\ft{21}{16}\,\Gamma_{ {a}}\chi_{R/L}\,
(\overline{\lambda}_{L/R}\,\Gamma^{ {a}}\lambda_{L/R}) \\
&& \mp\ft{1}{1280} \, \Gamma_{ {a_1}\dots
 {a_5}}\chi_{R/L}\,
(\overline{\lambda}_{L/R}\,\Gamma^{ {a_1}\dots
 {a_5}}\lambda_{L/R}\\
\mathcal{S}_{R/L}\lambda_{L/R} & = & 0\\
\hline
\mathcal{S}_R \, w_+ & = & \mathbf{d}_+ \\
\mathcal{S}_L \, w_+& = & 0 \\
\mathcal{S}_R \, w_- & = &0 \\
\mathcal{S}_L \, w_-& = & \mathbf{d}_-\\
\mathcal{S}_R \, \mathbf{d}_+ & = & 2{\rm i}\,\mathbf{\Gamma}_{ {a}} \, \Pi^{ {a}}_+\,\lambda_R\\
\mathcal{S}_L \, \mathbf{d}_- & = & -2{\rm i}\,\mathbf{\Gamma}_{ {a}} \, \Pi^{a}_-\,\lambda_L \\
\mathcal{S}_{L/R}\, \mathbf{d}_\pm &=& 0\\
\hline
\end{array}$
\end{center}
\end{table}
\begin{table}[!htb]
  \caption{\textbf{BRST algebra: }{\it In this table we display the BRST transformations of the
  composite fields, namely of the various field strenghts.}}\label{tbl:composite}
\begin{center}
{\small $
\begin{array}{|rcl|}
\hline
\mathcal{S}_{L/R}\,\mathcal{G}_{ {ab}}&=&e^{-\varphi}\,\left(\pm\overline{\lambda}_{L/R}\,\rho^{R/L}_{ {ab}}-\ft
32\,i\,f_{[ {a}}\,\overline{\chi}_{L/R}\,\Gamma_{ {b}]}\,\lambda_{L/R}+\ft
32\,i\,\mathcal{D}_{[ {a}}\,\overline{\chi}_{L/R}\,\Gamma_{ {b}]}\,\lambda_{L/R}\right.\\&&\left.+\ft
32\,i\,\overline{\chi}_{L/R}\,\Gamma_{[ {a}}\,\mathcal{L}^{(even)}_{ {b}]\pm}\lambda_{L/R}+\ft
32\,i\,\overline{\chi}_{R/L}\,\Gamma_{[ {a}}\,\mathcal{L}^{(odd)}_{ {b}]\pm}\lambda_{L/R}\right)\\
\mathcal{S}_{L/R}\,\mathcal{G}_{ {abcd}}&=&e^{-\varphi}\,\left(\overline{\lambda}_{L/R}\,\Gamma_{[ {ab}}\rho^{R/L}_{ {cd}]}
\pm\ft
i2\,f_{[ {a}}\,\overline{\chi}_{L/R}\,\Gamma_{ {bcd}]}\,\lambda_{L/R}\mp\ft
i2\,\mathcal{D}_{[ {a}}\,\overline{\chi}_{L/R}\,\Gamma_{ {bcd}]}\,\lambda_{L/R}\right.\\&&\left.\mp\ft
i2\,\overline{\chi}_{L/R}\,\Gamma_{[ {abc}}\,\mathcal{L}^{(even)}_{ {d}]\pm}\lambda_{L/R}\pm\ft
i2\,\overline{\chi}_{R/L}\,\Gamma_{[ {abc}}\,\mathcal{L}^{(odd)}_{ {d}]\pm}\lambda_{L/R}-\ft
32\,i\,\mathcal{H}_{[ {abc}}\,
\overline{\chi}_{L/R}\,\Gamma_{ {d}]}\,\lambda_{L/R}\right)\\
\mathcal{S}_{L/R}\,\mathcal{H}_{ {abc}}&=& \mp 2
\,i\,\overline{\lambda}_{L/R}\,\Gamma_{[ {a}}\,\rho^{L/R}_{ {bc}]}\\
\mathcal{S}_{L/R}\,\mathcal{D}_{ {a}}\chi_{L/R}&=&-\ft
14\,(\overline{\lambda}_{L/R}\,\Theta_{ {cd}, {a}|R/L})\,\Gamma^{ {cd}}\,\chi_{L/R}+
\left[\mathcal{D}_{ {a}}\mathcal{N}^{(even)}_\pm-(\mathcal{N}\,\mathcal{L}_{ {a}})^{(even)}_\pm\right]\,\lambda_{L/R}\\
\mathcal{S}_{L/R}\,\mathcal{D}_{ {a}}\chi_{R/L}&=&-\ft
14\,(\overline{\lambda}_{L/R}\,\Theta_{ {cd}, {a}|R/L})\,\Gamma^{ {cd}}\,\chi_{R/L}+
\left[\mathcal{D}_{ {a}}\mathcal{N}^{(odd)}_\pm-(\mathcal{N}\,\mathcal{L}_{ {a}})^{(odd)}_\pm\right]\,\lambda_{L/R}\\
\mathcal{S}_{L/R}\,\rho_{ {ab}}^{L/R}&=&\Upsilon^{(even)}_{ {ab}\,\pm}\,\lambda_{L/R}-\ft
14\,
R_{ {cd}, {ab}}\,\Gamma^{ {ab}}\,\lambda_{L/R}+2\,\mathcal{P}_{L/R}[\lambda_{L/R}]\,\rho_{ {ab}}^{L/R}\\
\mathcal{S}_{L/R}\,\rho_{ {ab}}^{R/L}&=&\Upsilon^{(odd)}_{ {ab}\,\pm}\,\lambda_{L/R}\\
\hline
\end{array}$}
\end{center}
\end{table}

\begin{table}[!htb]
  \caption{\textbf{Pure Spinor Action:}{\it In this table we display the complete form of the pure spinor action for tyep IIA superstring in a general background. In the formulas below $ \mathcal{S}_{L}  \mathcal{M}_-$ and $mathcal{S}_{R}\mathcal{M}_-$ are given in  (\ref{smm}) and
  $\mathcal{S}_{R} \mathcal{S}_{L}  \mathcal{M}_- $ is given in (\ref{spp}) for $\chi=0$
    } }\label{tbl:act}
\begin{center}
{$
\begin{array}{|rcl|}
\hline
\mathcal{A} &=& \mathcal{A}_{GS} +    \mathcal{A}_{gf}^{\rm IIA} \\
\mathcal{A}_{GS} & = &
{\displaystyle \int}  \left(\Pi^{ {a}}_+ \,
  V^{ {b}} \, \eta_{ {ab}} \, \wedge \, e^+ \, - \,
  \Pi^{ {a}}_- \,
  V^{ {b}} \, \eta_{ {ab}} \, \wedge \, e^-
  \, + \, \ft 12 \Pi^{ {a}}_i\, \Pi^{ {b}}_j
  \, \eta^{ij}\, \eta_{ {ab}} \, e^+ \, \wedge \, e^- \, +
   \ft 12 \,  \mathbf{B}^{[2]}\right)\\
 \mathcal{A}_{gf}^{\rm IIA}  &=& {\displaystyle\int}
 \Big(  \overline{\mathbf d}_+ \, \psi_R \, \wedge \, e^+ +
\overline{\mathbf d}_-  \, \psi_L \, \wedge \, e^-   +  \frac{\rm i}{2}
\overline{\mathbf d}_+  \,  \mathcal{M}_-  \, {\mathbf  d}_- \nonumber \\
&-& \overline{w}_+ \left(\mathcal{S}_{R}\psi_R\right) \, \wedge \, e^+
- \overline{w}_- \, \left(\mathcal{S}_{L}  \psi_L\right) \, \wedge \, e^-  \nonumber \\
&-&  \frac{\rm i}{2} \,  \overline{w}_+  \left( \mathcal{S}_{R}
\mathcal{M}_-  \right) {\mathbf  d}_- + \frac{\rm i}{2} \,
\overline{\mathbf d}_+  \left( \mathcal{S}_{L}  \mathcal{M}_-
\right) {w}_- - \frac{\rm i}{2} \,  \overline{w}_+  \left(
\mathcal{S}_{R} \mathcal{S}_{L}  \mathcal{M}_-  \Big) {w}_-
\right)\,. \\
\hline
\end{array}$
}
\end{center}
\end{table}

\eject
\newpage
\vfill

\section{Derivation of type IIA supergravity in the string frame}
\label{reduczia}

The derivation of type IIA supergravity was done in two steps. In the first one, we
started from the $D=11$ supergravity FDA and from its rheonomic parametrization and
we reduce them on a circle. Next, we perform a Weyl rescaling and gravitino field redefinition
to go to the string frame. In the second step, we derived the rheonomic parametrization directly
by solving the Bianchi identities in the $D=10$ in the string frame. Here, we just sketch such a
derivation.

\subsection{The D=11 FDA}
We start from the FDA of $M$-theory whose complete set of curvatures is given below \cite{fredauria11,comments}:
\begin{eqnarray}
{\mathbf{T}}^{\overline{a}} & = & \mathcal{D}\mathbf{V}^{\overline{a}}
- {\rm i} \ft 12 \, \overline{\Psi} \, \wedge \, \Gamma^{\overline{a}} \, \Psi \label{11torsion}\\
{\mathbf{R}}^{\overline{ab}} & = & d\mathbf{\Omega}^{\overline{ab}} - \Omega^{\overline{ac}} \, \wedge \, \Omega^{\overline{cb}}
\label{11riemann}\\
\widehat{\rho} & = & \mathcal{D}\Psi \equiv d \Psi - \ft 14 \, \Omega^{\overline{ab}} \, \wedge \, \Gamma_{\overline{ab}} \, \Psi\label{11rho}\\
\mathbb{F}^{{[4]}} & = & d\mathbb{A}^{{[3]}} - \ft 12\, \overline{\Psi} \, \wedge \, \Gamma_{\overline{ab}} \, \Psi \,
\wedge \, \mathbf{V}^{\overline{a}} \wedge \mathbf{V}^{\overline{b}} \label{11F4}\\
\mathbb{F}^{{[7]}} & = & d\mathbb{A}^{{[6]}} -15 \, \mathbb{F}^{{[4]}} \, \wedge \,  \mathbb{A}^{{[3]}} - \ft {15}{2} \,
\, \mathbf{V}^{\overline{a}}\wedge \mathbf{V}^{\overline{b}} \, \wedge \, {\bar \Psi}
\wedge \, \Gamma_{\overline{ab}} \, \Psi
\, \wedge \, \mathbb{A}^{{[3]}} \nonumber\\
\null & \null & - {\rm i}\, \ft {1}{2} \, \overline{\Psi} \, \wedge \, \Gamma_{\overline{a_1 \dots a_5}} \, \Psi \,
\wedge \, \mathbf{V}^{\overline{a_1}} \wedge \dots \wedge \mathbf{V}^{\overline{a_5}}
\label{11F7}
\end{eqnarray}
In the above equations $\mathbf{V}^{\overline{a}}$ and
$\mathbf{\Omega}^{\overline{ab}}$ are respectively the $11D$ vielbein
and spin connection, $\Psi$ is the $11D$ gravitino, namely a Majorana
spinor valued $1$-form of fermionic type with $32$--components, while
$\mathbb{A}^{{[3]}}$ and $\mathbb{A}^{{[6]}}$ are a bosonic $3$-form
and a bosonic $6$-form respectively. Equations
(\ref{11torsion},\ref{11riemann},\ref{11rho}) define the curvatures
of the $11D$ superPoincar\'e algebra. According to Sullivan's second
theorem the $3$-form $\mathbb{A}^{{[3]}}$ corresponds to the first
FDA extension of this latter generated by a degree $4$ cohomology
class, while the $6$-form $\mathbb{A}^{{[6]}}$ corresponds to a
further extension of the FDA generated by a degree $7$ cohomology
class of the first extension.
\par
The rheonomic parametrization of the M-theory curvatures is the
following one:
\begin{eqnarray}
\mathbf{T}^{\overline{a}} & = & 0 \label{torsequa}\\
\mathbb{F}^{{[4]}} & = & {F}_{\overline{a_1\dots a_4}} \, \mathbf{V}^{\overline{a_1}} \, \wedge \dots
\wedge \, \mathbf{V}^{\overline{a_4}} \label{f4equa}\\
\mathbb{F}^{{[7]}} & = & \ft {1}{84} {F}^{\overline{a_1\dots a_4}} \, \mathbf{V}^{\overline{b_1}} \, \wedge \dots \wedge \,
\mathbf{V}^{\overline{b_7}} \, \epsilon_{\overline{a_1 \dots a_4 b_1 \dots b_7}} \label{f7equa}\\
\widehat{\rho} & = & \rho_{\overline{a_1a_2}} \,\mathbf{V}^{\overline{a_1}} \, \wedge \, \mathbf{V}^{\overline{a_2}} + {\rm i} \ft 13 \,
\left(\Gamma^{\overline{a_1a_2 a_3}} \Psi \, \wedge \, \mathbf{V}^{\overline{a_4}} - \ft 1 8
\Gamma^{\overline{a_1\dots a_4 m}}\, \Psi \, \wedge \, \mathbf{V}^{\overline{m}}
\right) \, {F}_{\overline{a_1 \dots a_4}} \label{rhoequa}\\
\mathbf{R}^{\overline{ab}} & = & {R}^{\overline{ab}}_{\phantom{ab}\overline{cd}} \, \mathbf{V}^{\overline{c}} \, \wedge \,
\mathbf{V}^{\overline{d}}
+ {\rm i} \, \overline{\rho}_{\overline{mn}} \, \left( \ft 12 \Gamma^{\overline{abmnc}} - \ft 2 9
\Gamma^{\overline{mn}[\overline{a}}\, \delta^{\overline{b}]\overline{c}} + 2 \,
\Gamma^{\overline{ab}[\overline{m}} \, \delta^{\overline{n}]\overline{c}}\right) \,
\Psi \wedge \mathbf{V}_{\overline{c}}\nonumber\\
 & &+\overline{\Psi} \wedge \, \Gamma_{\overline{mn}} \, \Psi \, {F}^{\overline{mnab}} +
 \ft 1{24} \overline{\Psi} \wedge \,
 \Gamma^{\overline{abc_1 \dots c_4}} \, \Psi \, {F}_{\overline{c_1 \dots c_4}}
\label{rheoFDA}
\end{eqnarray}
and it implies the following field equations on the space-time
components:
\begin{eqnarray}
0 & = & \mathcal{D}_{\overline{m}} F^{\overline{mc_1 c_2 c_3}} \, +
\, \ft 1{96} \, \epsilon^{\overline{c_1c_2c_3 a_1 a_8}} \, F_{\overline{a_1 \dots a_4}}
\, F_{\overline{a_5 \dots a_8}}  \nonumber\\
0 & = & \Gamma^{\overline{abc}} \, \rho_{\overline{bc}} \nonumber\\
R^{\overline{am}}_{\phantom{\overline{bm}}\overline{cm}} & = & 6
\, F^{\overline{ac_1c_2c_3}} \,F^{\overline{bc_1c_2c_3}} -
\, \ft 12 \, \delta^{\overline{a}}_{\overline{b}} \, F^{\overline{c_1 \dots c_4}} \,F^{\overline{c_1 \dots c_4}}
\label{fieldeque}
\end{eqnarray}
In all the above equations the overlined latin indices run on eleven
values:
\begin{equation}
  \overline{a_1,a_2},\dots \, = \, 0,1,2,\dots,10
\label{indicivalori}
\end{equation}
\subsection{The type IIA FDA from circle reduction}
The $D=10$ Free Differential algebra defined in eqs.~(\ref{2acurva})-(\ref{Gcurva}) and its
rheonomic parametrization in eqs.~({\ref{nullatorsioSF}})-(\ref{dechiparaSF})
can now be obtained by
dimensional reduction on a circle $\mathbb{S}^1$ of the algebraic
structure described in the previous subsection.
\par
Explicitly, the Kaluza-Klein ansatz realting the $D=11$ with the $D=10$ items is the following:
\begin{eqnarray}
\mathbf{V}^{ {a}} & = & \exp{[- \ft 13 {\varphi}]}V^{ {a}} \nonumber\\
\mathbf{V}^{11} & = & \exp \left[ \ft 23 \, \varphi \right] \, \left(
d\theta \, + \, \mathbf{A}^{[1]} \right) \nonumber\\
\mathbf{A}^{[3]} & = & \mathbf{C}^{[3]} \, + \, \mathbf{B}^{[2]} \,
\wedge \, \left(
d\theta \, + \, \mathbf{A}^{[1]} \right) \nonumber\\
\Psi & = & \exp{[-\ft 16 \varphi]} \left( \psi_L \, + \, \psi_R\right) \, + \, \left(\chi_L \, + \, \chi_R
\right)\, \exp \, \left[\ft 56 \, \varphi  \right] \, \left(
d\theta \, + \, \mathbf{A}^{[1]} \right) \nonumber \\
&-& \frac{i}{2} \exp{[-\ft 16 \varphi ]} \Gamma_{  r}  \left(\chi_L \, - \, \chi_R
\right)\, V^{  r}\,,
\label{fundansatz}
\end{eqnarray}
where $\theta$ is the coordinate on the circle.
\par
Inserting this ansatz in the $D=11$ curvatures and redefining the $D=10$  spin connection in such a way
that the $D=10$ torsion is zero, we obtain eqs.~(\ref{2acurva})-(\ref{Gcurva}) and
({\ref{nullatorsioSF}})-(\ref{dechiparaSF}). Furthermore, from the above KK ansatz inserted in the
field equations (\ref{fieldeque}), we get the bosonic field equations of type IIA supergravity in sec. 2.4.


\section{Conventions}
In this appendix we collect all the relevant conventions for the Gamma matrix algebra
utilized in the
main text
\begin{eqnarray}
  \left\{\Gamma_{  a} \, , \, \Gamma_{  b} \right\} & = & 2\, \eta_{ab}
  \quad ; \quad {  a},{  b} = 0,\dots,9
  \label{dirac10}
\end{eqnarray}
\begin{equation}
  \eta_{  ab}  =  \mbox{diag}\{ +,-,-,-,-,-,-,-,-,-\}
\label{etaD10}
\end{equation}
\begin{equation}
\Gamma_0^\dagger = \Gamma_0\,, ~~~~
\Gamma_{11}^\dagger = \Gamma_{11}\,, ~~~~
\Gamma_{11} \psi_{L/R} = \pm \psi_{L/R}\,.
\end{equation}
We define the charge conjiugation matrix ${\mathcal C} \Gamma_{  a} {\mathcal C}^{-1} = - \Gamma_{  a}^T$. Due to these definitions ${\mathcal C} \Gamma_{  a}$,
${\mathcal C} \Gamma_{11} \, {\mathcal C}  \Gamma_{  {ab}}$,${\mathcal C}  \Gamma_{11 {bcde}}, {\mathcal C}  \Gamma_{ {abcde}}$ are symmetric and
 ${\mathcal C} , C\Gamma_{11 {ab}}, {\mathcal C} \Gamma_{  {abc}}, \Gamma_{11  {abc}}, {\mathcal C}  \Gamma_{ {abcd}}$ are antisymmetric.

\end{document}